\PassOptionsToPackage{dvipsnames}{xcolor}
\documentclass[manuscript]{acmart}

\AtBeginDocument{%
  }

\setcopyright{acmlicensed}
\copyrightyear{2024}
\acmYear{2024}
\acmDOI{XXXXXXX.XXXXXXX}

\acmConference[Preprint]{}{Conference acronym 'XX}{2024}

\acmISBN{978-1-4503-XXXX-X/18/06}

\usepackage{multirow}
\usepackage{booktabs}
\usepackage{listings}
\usepackage{fontawesome5}
\usepackage[dvipsnames]{xcolor}
\usepackage{soul}
\usepackage{tcolorbox}
\usepackage{xspace}
\usepackage{tikz}

\definecolor{explainpink}{rgb}{1, 0.6, 1}
\definecolor{explainblue}{rgb}{0.4, 0.4, 1}
\definecolor{explainred}{rgb}{1, 0.4, 0.4}
\definecolor{explaingreen}{rgb}{0.59608, 0.83137, 0.09412}

\newcommand{\SYSTEM}{\textsc{IntelliExplain}\xspace}

\newcommand{\highlight}[2]{%
    \begingroup
    \definecolor{hlcolor}{HTML}{#1}%
    \sethlcolor{hlcolor}%
    \hl{#2}%
    \endgroup
}
\begin{document}

\title{\SYSTEM: Enhancing Conversational Code Generation for Non-Professional Programmers}


\author{Hao Yan}
\affiliation{%
  \institution{George Mason University}
  \city{Fairfax}
  \state{VA}
  \country{USA}
}
\email{hyan5@gmu.edu}

\author{Thomas D. LaToza}
\affiliation{%
  \institution{George Mason University}
  \city{Fairfax}
  \state{VA}
  \country{USA}
}
\email{tlatoza@gmu.edu}

\author{Ziyu Yao}
\affiliation{%
  \institution{George Mason University}
  \city{Fairfax}
  \state{VA}
  \country{USA}
}
\email{ziyuyao@gmu.edu}

\renewcommand{\shortauthors}{Yan, Latoza, and Yao}

\begin{abstract}
Chat LLMs such as GPT-3.5-turbo and GPT-4 have shown promise in assisting humans in coding, particularly by enabling them to conversationally provide feedback. However, current approaches assume users have expert debugging skills, limiting accessibility for non-professional programmers. In this paper, we first explore Chat LLMs' limitations in assisting non-professional programmers with coding. Through a formative study, we identify two key elements affecting their experience: the way a Chat LLM explains its generated code and the structure of human-LLM interaction. We then propose \SYSTEM, a new conversational code generation framework with enhanced code explanations and a structured interaction paradigm, which enforces both better code understanding and a more effective feedback loop. In two programming tasks (SQL and Python), \SYSTEM yields significantly higher success rates and reduces task time compared to the vanilla Chat LLM. We also identify several opportunities that remain in effectively offering a chat-based programming experience for non-professional programmers.
\end{abstract}


\begin{teaserfigure}
  \includegraphics[width=\textwidth]{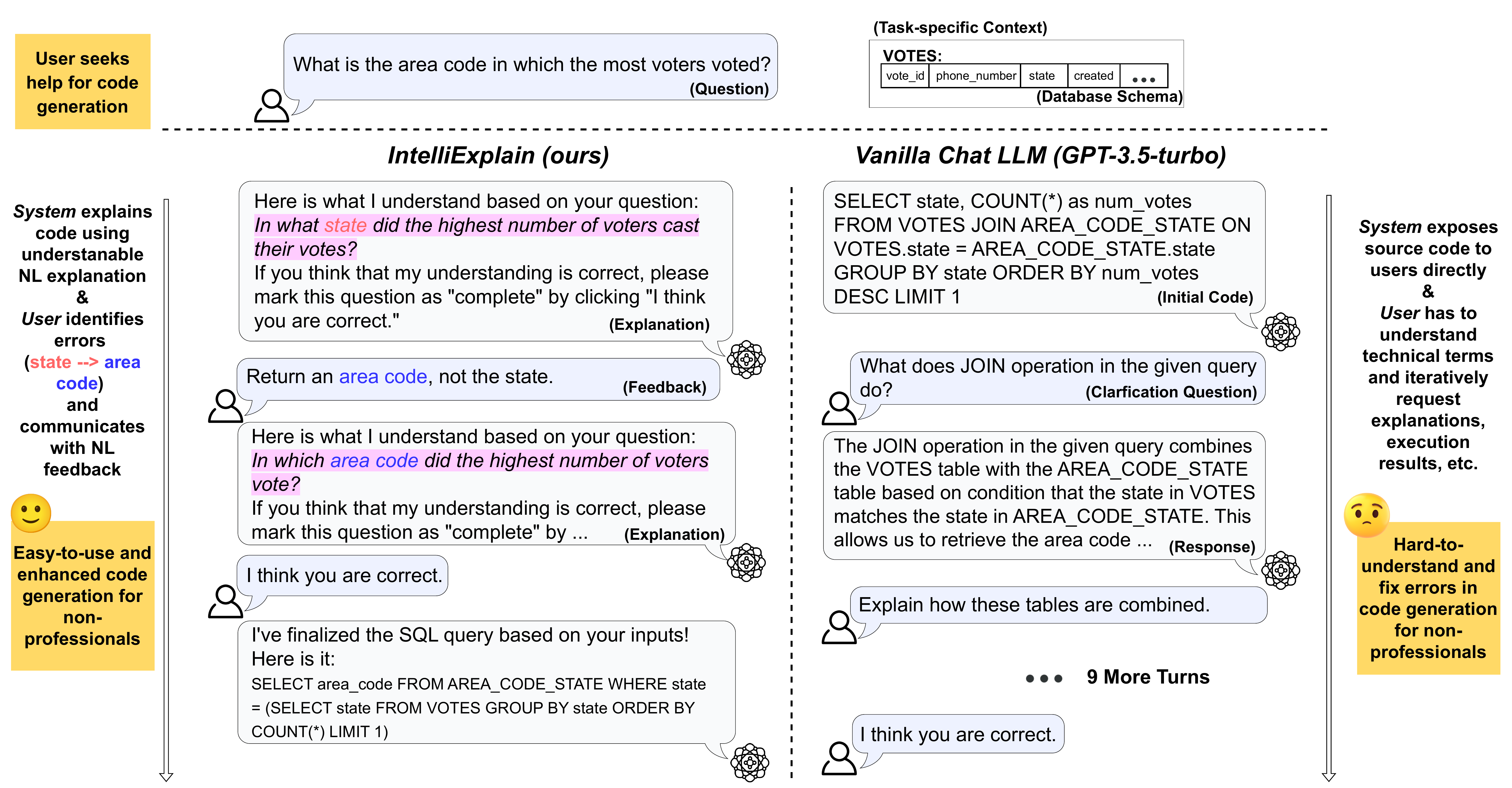}
  \caption{
    \SYSTEM\ enables non-professional programmers to write code in natural language (NL) without requiring direct interaction with code. A user starts with a question in NL, accompanied by relevant context (top). \SYSTEM\ then generates source code and confirms its understanding of the question by presenting an NL explanation (in \textcolor{explainpink}{\faSquare}) to the user. When this understanding is incorrect, the user can provide corrective feedback in NL and instruct the system.}
  \label{fig:overview}
\end{teaserfigure}

\received{20 February 2007}
\received[revised]{12 March 2009}
\received[accepted]{5 June 2009}

\maketitle

\section{Introduction}
The field of AI-powered code generation has witnessed a significant paradigm shift with the emergence of Large Language Models (LLMs) such as Codex~\cite{chen2021evaluating}, Code Llama~\cite{roziere2023code}, StarCoder~\cite{li2023starcoder}, and CodeT5~\cite{wang2021codet5}. Unlike prior approaches that often involved labor-intensive data collection and annotation efforts, current LLMs can learn directly from few-shot task demonstrations fed in their prompt context (called ``few-shot, in-context learning'')~\cite{liu2023pre}. Given a task description provided in natural language (NL), or sometimes an incomplete code snippet, these LLMs generate or complete the code by autoregressively generating a sequence of code tokens. The advent of Chat-based Large Language Models (Chat LLMs) like GPT-3.5\&4 \cite{openai:gpt, achiam2023gpt}, Claude~\cite{claude}, and Gemini~\cite{gemini} further support code generation by offering real-time, interactive assistance. These models are designed for conversational use and allow users to actively engage in LLM's decision-making process and provide real-time feedback (called ``conversational code generation''). This conversation feature of Chat LLMs greatly enhances the code quality and better ensures that the code fits user goals. As a result, Chat LLMs have become the standard for real-time coding assistance.

However, how to leverage Chat LLMs' interactive features in assisting \emph{non-professional programmers} to write code remains a challenge. Non-professional programmers encompass both novice and end-user programmers. Novice programmers are individuals who are new to programming and have limited experience or knowledge in writing code. In contrast, end-user programmers are individuals who may not have formal training in coding but use programming interfaces to automate tasks, develop scripts, or modify existing software applications for personal or professional use. Prior works have explored how programmers interact with LLM-based code assistants from a usability perspective in a non-conversational setting. For example, \citet{vaithilingam2022expectation} and \citet{barke2023grounded} systematically examined the usability of Github Copilot~\cite{github:copilot} in assisting experienced programmers. They found that the Copilot is effective at providing a starting point but requires further professional debugging to ensure correctness. \citet{kazemitabaar2023studying} investigated how GitHub Copilot assisted novice programmers in an introductory programming class and found that it significantly increased coding performance during the learning phase without decreasing performance on later manual code modification tasks.

In a conversational code generation scenario, a user typically interacts with a Chat LLM by posing programming questions and optionally providing input-output samples to specify requirements. When errors are identified or when the generated code fails to meet the specified requirements, users follow up with either clarifications or corrective feedback pinpointing the errors and prompting the model to refine the code solution. \citet{ross2023programmer} showed that conversational assistants could provide valuable assistance to software engineers by enabling them to ask follow-up questions that depend upon their conversational and code contexts. \citet{khojah2024beyond} highlighted the usage of ChatGPT in generating high-level guidance. However, these explorations have focused on only professional programmers, yet how a Chat LLM-based code assistant can help non-professional programmers is underexplored.

Despite the simplicity of the conversational paradigm in code generation, the difficulty in understanding the generated code and accurately pinpointing and articulating errors within the code makes it challenging for users, especially non-professional programmers, to provide meaningful corrective feedback. To address the problem, prior research has conducted extensive exploration of the formats of user feedback for code error correction~\cite{elgohary-etal-2020-speak, wang2022compilable, yang2023intercode, wang2023leti, chen2023improving}, but there was only limited focus on developing solutions specifically for non-professional programmers with Chat LLMs. One recent work conducted by \citet{prather2023s} revealed that novice programmers often struggle to understand LLM-generated code and likely accept incorrect code suggestions, which also highlights the need for forming a better understanding of how non-professional programmers interact with (Chat) LLMs and proposing a better conversational code generation framework for them. 

In this work, we first provide a systematic study toward understanding how non-professional programmers interact with the vanilla Chat LLM in a conversational code generation scenario. Specifically, we performed a user study with a group of participants with no or only limited expertise in programming, tasking them with 10 programming tasks in SQL and another 10 in Python and observing their interaction patterns with a vanilla Chat LLM. We measured their rate of successfully completing the coding tasks and analyzed how their interaction patterns had various impacts on the success rate. Our analysis identified two critical elements — \emph{code explanations} and \emph{structure of human-LLM interaction} — as essential for allowing the user to understand the model-generated code and provide effective feedback for error correction. Nonetheless, our study showed that a vanilla Chat LLM, due to the lack of a carefully designed explanation generation method and the similar lack of a well-structured human-LLM interaction procedure, is not able to assist non-professional programmers with sufficient efficacy. In fact, participants reported frustration from interacting with the vanilla Chat LLM.

To address these issues, we introduce \SYSTEM, a new conversational framework equipped with enhanced explanations and a structured human-LLM interaction paradigm to assist non-professional programmers in comprehending and debugging code. 
\SYSTEM follows a well-structured interaction procedure with humans. It first provides an NL explanation of its generated source code to the user, then prompts the user to identify issues and give NL feedback based on the explanation, and finally refines the code solution according to the user feedback. This approach allows non-professional programmers to write code using natural language, without requiring professional programming knowledge or direct interaction with the source code (Figure~\ref{fig:overview}). The key insight of \SYSTEM lies in its use of an enhanced NL explanation of code, which presents a more accessible version of the source code while offering users with a clear understanding of code logic. Its structured interaction procedure helps users to provide more effective feedback and eventually yields a higher success rate in the coding tasks.

To investigate the effectiveness of \SYSTEM, we conducted a second user study, reusing the setting from our study with the vanilla Chat LLM. Our results indicate that participants using \SYSTEM achieved 11.53\% and 25.31\% higher success rates while requiring less time to write correct code compared to those using the vanilla Chat LLM on both tasks respectively. Even participants with no prior programming experience were able to write and debug code solely by relying on our enhanced NL explanations and following the structured interaction paradigm we designed. The results confirmed the importance of the two elements we discovered and demonstrated the effectiveness and efficiency of \SYSTEM. Finally, we see that even with the advancement we have made with \SYSTEM, further room remains for future improvement. We identified additional challenges within conversational code generation that hinder non-professional programmers and highlighted potential future research directions in this area. To summarize:

\begin{itemize}
    \item{We systematically studied how non-professional programmers interact with the vanilla Chat LLM in conversational code generation and identified key elements (i.e., code explanations and the human-LLM interaction structure) that affected its efficacy.}
    \item{We introduced \SYSTEM, a conversational framework based on a novel structured human-LLM interaction paradigm that incorporates our enhanced NL explanations for conversational code generation.}
    \item{Our user study shows that \SYSTEM helps non-professional programmers, including those with no prior experience, to write and debug code more effectively and efficiently.}
    \item We included a thorough analysis of two user studies and discussed the potential and challenges for future researchers to continue the effort of enhancing Chat LLM-based conversational code generation for non-professional programmers.
\end{itemize}
\section{Related Work}
\subsection{Interactive and Conversational Code Generation}
Interactive code generation, where users interactively work with a tool or a system for code generation, has been a long-standing problem. One example is allowing users to interactively provide or annotate input-output examples in the scenario of programming by example (PBE). For example, \citet{zhang2020interactive} introduced mechanisms to augment user-provided examples. Users could either directly annotate desired and undesired parts of generated regular expressions or identify counterexamples synthesized by the model for further code refinement. \citet{drosos2020wrex} developed Wrex, a Jupyter Notebook extension using a programming-by-example environment for interactive data transformations. It allows users to provide transformation examples via an interactive grid, where Wrex then generates and inserts readable code into the notebook for immediate application and visualization, streamlining the data preprocessing. However, even proficient programmers cannot provide representative examples that cover as many practical situations as possible. Consequently, while the provided or augmented examples may match the expected output, the output code may still produce undesired behavior in unseen cases.

Research on conversational code generation has been an active topic even before the recent popularization of Chat LLMs \cite{chaurasia-mooney-2017-dialog, su2018natural, labutov-etal-2018-learning, yao2019interactive, staniek2021error, yao-etal-2020-imitation, li-etal-2020-mean, elgohary-etal-2020-speak, mo-etal-2022-towards}. A conversational code generation system typically consists of three components: a code generator, a mechanism to identify and request user feedback on the predicted code, and an error correction model to refine the code based on user feedback. Studies conducted by \citet{gur-etal-2018-dialsql} and \citet{yao-etal-2019-model} approached this by explaining components in a generated SQL code, and if any component was wrong, users were prompted to select the correct components from a shortlist as feedback. Another approach, proposed by \citet{li-etal-2020-mean}, identified uncertain tokens in the user's NL commands and sought user choices for paraphrases to enhance clarity. However, the multi-choice feedback type adopted by these prior approaches, while showing promise in the text-to-SQL task they focused on, exhibited limitations in terms of user-friendliness, efficiency, and generalizability to more complex programming languages. In particular, users could only passively respond to system-presented choices, posing challenges in facilitating a more dynamic and user-centric interaction. To address this issue, free-form NL feedback has been introduced \cite{labutov-etal-2018-learning, elgohary-etal-2020-speak}. \citet{elgohary-etal-2020-speak} demonstrated the effectiveness in correcting code errors via NL feedback and annotated the SPLASH dataset to benchmark automatic error correction. Their subsequent work NL-Edit \cite{elgohary-etal-2021-nl} further fine-tuned a model to convert the NL feedback into actionable code edits for error correction. However, their work was conducted to correct errors made by Seq2Struct \cite{shin2019encoding}, a much weaker code generator than the current (Chat) LLMs. Therefore, the effectiveness of their error correction model may not generalize to errors made by the more advanced LLMs, and the success rate reported in their work is not comparable with the results in our study. In addition, given their need for annotating edit data to train the error correction model, their model was task- and programming language-specific, whereas in our work, we aim for a conversational system that can generalize across tasks and programming languages.

As LLMs improve their conversation ability, several studies have explored Chat LLMs for code generation. \citet{champa2024chatgpt} conducted a quantitative analysis of 2,865 developer-ChatGPT conversations from the DevGPT~\cite{xiao2024devgpt} dataset, examining how developers use ChatGPT across 12 software task categories. The study found that the ChatGPT is most effective for tasks like software development management, optimization, and new feature implementation, but less efficient in areas such as environment setup, documentation, and code quality management. \citet{chopra2023conversational} systematically examined conversational assistants for data scientists and identified challenges such as contextual data retrieval, adapting generated code to local environments, and refining prompts. While other works have explored ChatGPT’s usability for code completion~\cite{sridhara2023chatgpt, rabbi2024ai} and debugging~\cite{ge2023empirical,surameery2023use}, few focus on its conversational role in assisting non-professional programmers from both aspects. The use of (Chat) LLMs for code generation in introductory programming classes has also gained popularity among both students and educators~\cite{kazemitabaar2023novices,kazemitabaar2023studying,prather2023s,sheese2024patterns,becker2023programming}. Unlike these studies, which focus on the educational setting and studying the impact of (Chat) LLMs on programming learning, our work is centered on whether and how these models can assist non-professional programmers for better task completion. However, we envision that the insights we discovered, such as the necessity of enhancing the explanations of model-generated code and having more structured human-LLM interaction, can generalize to the educational setting.

\subsection{Code Comprehension}
While LLM-based coding tools allow programmers with limited coding skills or domain knowledge to write code more easily, they also introduce challenges, as these programmers may struggle to understand and debug the generated code \cite{vaithilingam2022expectation, barke2023grounded}. These challenges have motivated the need for effective code comprehension support. In this space, \citet{nam2024using} developed a Visual Studio Code plugin, which enhances code comprehension by triggering AI explanations on selected code or providing detailed explanations as a response to user queries. \citet{yan2024ivie} introduced Ivie, a tool providing instant, in-situ AI explanations for generated code. Ivie integrated LLMs to display concise explanations next to code, from variables to entire blocks.
A lab study showed that Ivie improved code understanding and was a useful, low-distraction addition to programming assistants. \citet{leinonen2023comparing} found that while LLM-generated and student-generated code explanations are similar in length, LLMs' explanations are perceived as more accurate and easier to understand. \citet{sarsa2022automatic} examined the abilities of LLMs in generating programming exercises and code explanations, finding that most of the generated content is both novel and coherent. However, none of these studies comprehensively examined how code explanations impact the interaction experience of non-professional programmers using Chat LLMs for code generation. Our work thus complements existing research.
\section{User Study Design}\label{sec:user-study}
To understand the behaviors and challenges encountered by non-professional programmers in using Chat LLMs for programming, we conducted a formative study. All of the human subject studies involved in our work have received approval from the university's institutional review board (IRB). Participants worked on two coding tasks: Text-to-SQL and Python code generation. Both required translating natural language questions into executable code: SQL queries for Text-to-SQL and Python code for the Python task. Concentrating on understanding how non-professional programmers interact with a Chat LLM for debugging, we selected tasks for which the LLM writes incorrect code.

\subsection{Setup} \label{sec:setup}

\begin{table}[t!]
    \resizebox{0.9\textwidth}{!}{%
    \begin{tabular}{p{0.1\linewidth}p{0.5\linewidth}p{0.1\linewidth}p{0.2\linewidth}}
    \toprule
    & \multicolumn{3}{c}{\textbf{Text-to-SQL}} \\
    \midrule
    \multicolumn{1}{p{0.1\linewidth}|}{\textbf{Difficuty Level}} & \multicolumn{1}{p{0.5\linewidth}|}{\textbf{Sample Question \& Edits for SQL {\small (\textcolor{red}{\st{incorrect}} to \textcolor{ForestGreen}{correct})}}} & \multicolumn{1}{p{0.1\linewidth}|}{\textbf{Edits Count}} & \textbf{Syntax Complexity of Generated Code} \\
    \midrule
    \multicolumn{1}{p{0.1\linewidth}|}{\textbf{Easy}} & \multicolumn{1}{p{0.5\linewidth}|}{``What is the grade of each high schooler?'' \newline 
    {\small {SELECT \textcolor{red}{\st{ID,}} grade FROM Highschooler}}
    } & \multicolumn{1}{p{0.1\linewidth}|}{1-2 actions} & \multicolumn{1}{c}{-} \\
    \midrule
    \multicolumn{1}{p{0.1\linewidth}|}{\textbf{Medium}} & \multicolumn{1}{p{0.5\linewidth}|}{``Count the number of countries for which Spanish is the predominantly spoken language.'' \newline 
    {\small {SELECT COUNT(*)\textcolor{ForestGreen}{, MAX(Percentage)} FROM countrylanguage WHERE \textcolor{red}{\st{IsOfficial='T'}} \textcolor{ForestGreen}{LANGUAGE="Spanish"} \textcolor{ForestGreen}{GROUP BY CountryCode}}}
    } & \multicolumn{1}{p{0.1\linewidth}|}{3-5 actions} & \multicolumn{1}{c}{-} \\
    \midrule
    \multicolumn{1}{p{0.1\linewidth}|}{\textbf{Hard}} & \multicolumn{1}{p{0.5\linewidth}|}{``What is the area code in which the most voters voted?'' \newline 
    {\small SELECT \textcolor{red}{\st{state}} \textcolor{ForestGreen}{area\_code} FROM votes as T1 \textcolor{ForestGreen}{JOIN area\_code\_state AS T2 ON T1.state = T2.state}} GROUP BY \textcolor{red}{\st{state}} \textcolor{ForestGreen}{area\_code} ...
    } & \multicolumn{1}{p{0.1\linewidth}|}{\textgreater 5 actions} & \multicolumn{1}{c}{-} \\ 
    \midrule
    & \multicolumn{3}{c}{\textbf{Python Code Generation}} \\ 
    \midrule
    \multicolumn{1}{p{0.1\linewidth}|}{\textbf{Easy}} & \multicolumn{1}{p{0.5\linewidth}|}{``Write a function to round the given number to the nearest multiple of a specific number.''} & \multicolumn{1}{p{0.1\linewidth}|}{1-2 lines} & \multicolumn{1}{p{0.2\linewidth}}{{Basic math expressions, single for-loops, simple if-then conditions}} \\
    \midrule
    \multicolumn{1}{p{0.1\linewidth}|}{\textbf{Medium}} & \multicolumn{1}{p{0.5\linewidth}|}{``Write a python function to check whether the given number can be represented by product of two squares or not.''} & \multicolumn{1}{p{0.1\linewidth}|}{3-5 lines} & \multicolumn{1}{p{0.2\linewidth}}{{Nested for-loops, basic recursion}} \\
    \midrule
    \multicolumn{1}{p{0.1\linewidth}|}{\textbf{Hard}} & \multicolumn{1}{p{0.5\linewidth}|}{``Write a python function to find the last digit when factorial of a divides factorial of b.''} & \multicolumn{1}{p{0.1\linewidth}|}{Completely Rewrite} & \multicolumn{1}{p{0.2\linewidth}}{{Potentially with more complex recursions, deeper for loops, advanced data structures}} \\
    \bottomrule
    \end{tabular}%
    }
    \caption{Sample test questions for each difficulty level. The edits of text-to-SQL were calculated based on the number of actions needed to correct the errors. The edits of Python were manually counted based on the required revisions on the predicted code.
    }
    \vspace{-\baselineskip}
    \label{tab:sample_questions}
\end{table}

We employed GPT-3.5-turbo (version 0613) as our backend Chat LLM. For text-to-SQL, we utilized Spider~\cite{yu-etal-2018-spider}, a large-scale, complex, and cross-domain dataset, and for the Python code generation, we used MBPP~\cite{austin2021program}, which contains entry-level programming questions. We selected 10 questions from each dataset where GPT-3.5 exhibited errors in its initial few-shot code generation. To maximize the potential insights we could collect from the user study, we carefully categorized questions in each dataset into three difficulty levels --- easy, medium, and hard, and randomly selected questions from each category to form the 10 test questions used in the study. Examples of questions at different difficulty levels are shown in Table~\ref{tab:sample_questions}.

More specifically, for text-to-SQL, we initially followed the same criteria as \citet{yu-etal-2018-spider} and defined the question difficulty based on the number of components in the ground-truth SQL query, so that queries containing more SQL keywords (GROUP BY, ORDER BY, nested subqueries, etc) are considered to be harder. Under this criteria, we selected 4 easy, 3 medium, and 3 hard questions from Spider. However, our subsequent analysis revealed that a more complex ground-truth query does not necessarily imply a more challenging user interaction. For example, even for a very complex query, when the Chat LLM's initial prediction is almost correct, the required user interaction for code correction is minimal. Therefore, we redefined the difficulty level of SQL tasks based on the number of edits required to modify the initial prediction of the Chat LLM to the ground-truth query. The same idea was also adopted by \citet{elgohary-etal-2021-nl}. Under this new criteria, we defined difficulties as easy (within 2 edits), medium (3-5 edits), and hard (more than 5 edits), resulting in a new categorization of 4 easy, 2 medium, and 3 hard questions for the SQL tasks.

For Python code generation, we defined the difficulty level by the syntax complexity of predicted code and the edits of error correction. Specifically, {an easy question typically involves fewer lines of generated code and contains only basic logic, such as a single for loop, simple if-then conditions, and basic math expressions. Errors in easy questions are also simple to correct with minimal modifications (within 2 lines). A medium question includes slightly more complex logic, such as simple recursion or nested for loops, but the errors can be fixed within 5 lines without the need to rewrite the entire code. A hard question, at a minimum, shares similar logic complexity to a medium question but may involve more complex recursion, deeper nested loops, or advanced data structures. Errors in hard questions often require rewriting the entire code logic to ensure correctness.}

\subsection{Participants} \label{sec:participants} 
We recruited undergraduate students from various majors through recruitment flyers and email advertisements. Each applicant completed a demographic survey with items on their programming background and experience level. We selected participants for inclusion who were beginners in programming, including first-year computer science students who had only completed an introductory programming course with minimal practical experience as well as individuals from non-computer science majors, who had no prior programming experience but were familiar with basic mathematical logic and could benefit from programming in their work.
From 50 applicants, we recruited 22 participants who met our requirements, of which 20 completed the studies (10 for our formative study to be presented in Section~\ref{gpt_user_study} about vanilla GPT-3.5-turbo, and 10 for the study with our proposed framework, which will be introduced in Section~\ref{intelliexplain}). 18 reported no experience with databases or SQL queries, while the remaining 2 had taken university database classes but lacked practical experience. One participant had no prior experience with Python, while the other 19 had completed an introductory Python course.

\subsection{User Interface} \label{sec:user_interface}

\begin{figure*}[t!]
    \centering
    \includegraphics[width=\linewidth]{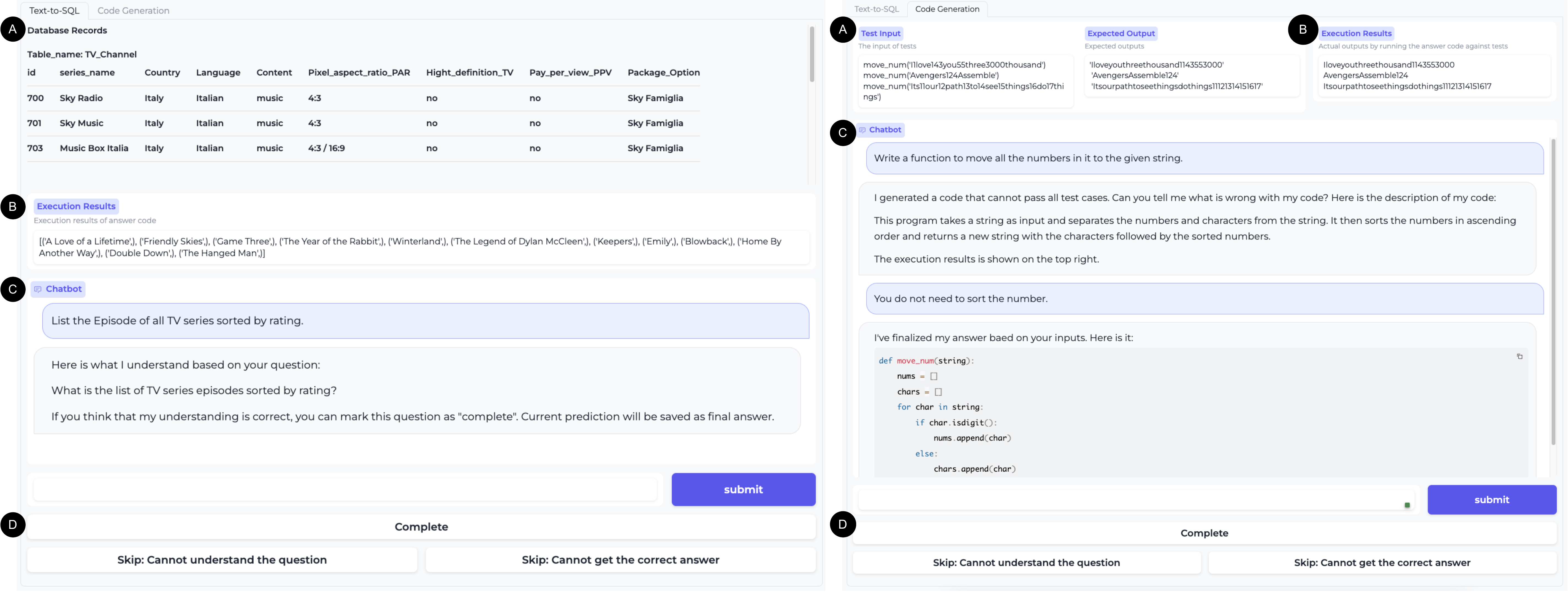}
    \caption{The user interface used in our user studies. For both tasks, the UI consists of 4 components: (A) Contextual information needed to answer the question (e.g., sample database records for Text-to-SQL and test cases and expected outputs for Python Code Generation); (B) (Only for \SYSTEM) Execution results, which are returned values by executing the predicted code against the database or test cases; (C) Chatbot interface, showing the conversation history between participants and the Chat LLM, including a text box for user input and a ``Submit'' button; and (D) Control panel, including two ``Skip'' buttons.}
    \label{fig:user_interface}
\end{figure*}

Understanding the question and its context—such as the database in text-to-SQL or test cases in Python code generation—is crucial for coding tasks. However, the default web-based interface of GPT-3.5-turbo (i.e., the ChatGPT web interface) does not support the inclusion of additional contextual information. To address the issue, we developed our own User Interface (UI) using Gradio (Figure~\ref{fig:user_interface}).\footnote{\hyperlink{https://www.gradio.app}{https://www.gradio.app}.} 
{For both tasks, the UI includes a panel displaying the contextual information of the task (A) --- for text-to-SQL, the information includes the database schema and three sample records for each table in the database; for Python code generation, the information consists of test cases and their expected outputs. The UI also contains a chatbot window showing the conversation history between the user and the Chat LLM, along with a text box for user input (C); and three buttons are included (D): a ``Complete'' button for participants to indicate that they believe the current prediction is correct and wish to end the interaction session, and two ``Skip'' buttons, one used when participants struggle to understand the question, and another used when participants feel that the Chat LLM cannot generate the correct answer after multiple attempts. In our second study, the UI additionally contains a panel displaying the execution results of the machine-generated code (B).}

\subsection{Study Procedure}
The user study comprised three phases: a warm-up session, a formal study, and a post-task interview. Recognizing that all participants entered the study with limited experience in the specified tasks and were unfamiliar with our UI, we initiated the study with a warm-up session. An experimenter provided an overview of the two tasks and introduced participants to the various functionalities embedded within our UI. Participants then actively engaged with the UI, tackling two warm-up questions for each task to foster familiarity and proficiency. Throughout this warm-up session, participants were encouraged to pose any questions related to the tasks or coding process, fostering a collaborative and informative environment. A fail-safe was included, so if participants found the training questions challenging or faced issues with the UI, they were guided to end their participation. Upon successful completion of the warm-up session, qualified participants then proceeded to the formal study, where they were tasked with independently solving all test questions. During the formal study, the experimenter played a passive role, intervening solely to address clarifying questions or resolve any technical issues encountered by the participants. This intentional shift allowed for an authentic assessment of the user study. To encourage efficient completion of the user study and to keep participants focused, we set a 5-minute time limit for each question. Participants were instructed to skip any question they could not solve within this time frame. After the study tasks, we conducted a semi-structured interview to explore participants' experiences and gather detailed feedback on conversational code generation. The interview questions covered various aspects, from overall experiences to specific components that assisted them in solving coding tasks or challenges they encountered during the formal study.

\subsection{Evaluation}
For evaluating the performance of a Chat LLM in assisting non-professional programmers, we report Success Rate (SR), defined as the percentage of success codes that, when executed, produce results matching the expected outputs. For text-to-SQL, SR is implemented with the official Execution Accuracy metric of Spider \cite{yu-etal-2018-spider}, which compares the execution results between the LLM-generated code and the ground-truth code against the database. For Python code generation, we execute the generated code to see whether the code can pass all test cases. In addition, we also report the average time spent per question (denoted as ``Avg. Time/Question''), which measures how efficiently a Chat LLM can assist participants in coding. For each metric, we calculate the micro-average among different questions and report the overall standard deviation.
\section{How Effectively Do Chat LLMs Assist Non-Professional Programmers in Coding?} \label{gpt_user_study}

\subsection{Task Performance}
Table \ref{tab:gpt_performance} presents the overall task performance results. We excluded from our analysis tasks in which the participants indicated difficulty in understanding the initial question and hence skipped it (i.e., ``Cannot understand the question''). In text-to-SQL, participants skipped a total of 10 tasks across 8 distinct questions, out of 100 overall (10 participants by 10 test questions). In the Python code generation tasks, participants together skipped 8 tasks across 6 distinct questions.

\begin{table}[t!]
    \resizebox{0.55\textwidth}{!}{%
    \begin{tabular}{l|cccc}
    \toprule
    \multirow{2}{*}{} & \multicolumn{4}{c}{\textbf{Success Rate (\%)}} \\ \cmidrule{2-5} 
    & \textbf{Easy}  & \textbf{Medium} & \multicolumn{1}{c||}{\textbf{Hard}} & \textbf{Overall} \\ 
    \midrule
    \textbf{Text-to-SQL} & 25.83 & 0.00    & \multicolumn{1}{c||}{0.00} & 10.33 ($SD=0.16$) \\
    \textbf{Python Code Gen} & 28.89 & 46.30  & \multicolumn{1}{c||}{3.33} & 26.44 ($SD=0.34$) \\ 
    \midrule
    & \multicolumn{4}{c}{\textbf{Ave. Time Spent/Question (s)}} \\ 
    \midrule
    \textbf{Text-to-SQL} & 174.95 & 87.75 & \multicolumn{1}{c||}{150.16} & 147.59 ($SD=57.53$)  \\
    \textbf{Python Code Gen} & 167.25 & 181.00 & \multicolumn{1}{c||}{200.13} & 181.24 ($SD=46.91$)  \\ 
    \midrule
    \end{tabular}%
    }
    \caption{Overall performance when participants interact with the vanilla Chat LLM (GPT-3.5-turbo). {We reported the micro-average success rate (in percentage) and time spent (in seconds) among questions. SD refers to standard deviation.}
    }
    \vspace{-\baselineskip}
    \label{tab:gpt_performance}
\end{table}

Overall, we found that, even with the assistance of a chat LLM in writing the code, successfully writing SQL was very challenging for non-professional programmers. Even for the questions we categorized as easy, there was a low SR of 25.83\%. None of the participants successfully completed any of the medium or hard questions: performance dropped to 0.0\%. 

Among the three difficulty levels of test questions, easy questions led to the longest interaction time, as these questions were most understandable by the participants, and as a result, participants had more engagement with the vanilla Chat LLM on them. These meaningful interactions yielded a non-zero SR as we described above. In contrast, participants had more difficulty in understanding the generated code from medium and hard questions, resulting in less time compared to their spent on the easy questions. We also noticed the time drop with medium questions. Our conjecture is that, while the participants were not able to engage deeply with the vanilla Chat LLM for both levels of questions, the rich information of hard questions gave them more room for interaction, hence the longer interaction time. However, no matter if they spent shorter or longer time on these questions, their SRs were both zeros.

The relatively higher SRs in Python tasks suggest that non-professional programmers may find Python's syntax and readability more accessible and that they are able to communicate more effectively with the Chat LLM. The time spent per question increases with task difficulty, showing that participants invested more time as the complexity of the tasks increased. An interesting observation here is that the SR of medium questions was even higher than that of easy questions. By looking into participants' conversations, we discovered that some participants requested a simulation of execution results for the generated code. As we will discuss in Section \ref{sec:gpt_ccd}, one limitation of the simulation process is that the Chat LLM may generate ``fake'' results to align with expected results but the faked result is not the actual output by running the code. The simpler logic of easy questions often leads to the simulation process only containing the faked result without sufficient intermediate logic. Participants based on the faked result could not provide accurate feedback for error correction. In contrast, medium questions involve more intermediate steps in the simulation process, and even if the final simulation result was faked, participants could spot errors by carefully examining these steps. Another issue was incorrect mental reasoning. Some participants attempted to provide feedback based on their own mental simulations or solutions for easy questions. In some cases, they miscalculated mathematical expressions or had incorrect logical reasoning, resulting in inaccurate feedback. However, this type of mental reasoning appeared less frequently in conversations involving medium questions.

Despite this, hard tasks still present a significant challenge. This reveals a gap in the ability of non-professional programmers to effectively communicate with Chat LLM for more complex coding tasks.

By analyzing the participants' conversations and post-task interviews, we identified two key challenges that significantly impacted participants' performance on the tasks:  \emph{code comprehension} and \emph{error correction}. In the next two subsections, we examine each of these challenges in detail.

\subsection{Code Comprehension} \label{sec:gpt_ccd}
Participants comprehended code by interact with the LLM to understand the source code and interpret the logic, structure, and semantic meaning of the code. 
Understanding the machine-generated code was important for participants to be able to correctly edit the code to fix issues which were present. Participants often struggled to read and interpret code and had to invent their own ways to prompt the LLM for code comprehension. We identified four distinct methods participants used to interact with the Chat LLM to understand the code (Table~\ref{tab:gpt_ccd_behavior}).

\begin{itemize}
    \item \textbf{Complete Explanation:} Across both tasks, over 85\% of participants requested a complete line-by-line explanation of the source code. This indicates that non-professional programmers often need comprehensive guidance to understand the overall logic and structure of the code. In reviewing the model's explanations, we found that they generally align with the generated code. However, we found two key drawbacks: \emph{(1) Excessive length and complexity:} The explanations provided were often too detailed, complex, and lengthy. Non-professional programmers, who may already find code comprehension difficult, may quickly lose patience or feel overwhelmed by the amount of information. As a result, they may skim through the explanations without fully understanding them. In our post-task interviews, 5 out of 10 participants reported experiencing this issue. \emph{(2) Use of technical language:} Even when the explanations were semantically correct, participants often struggled to understand them due to a lack of foundational programming knowledge. Post-task interviews revealed that many participants were unable to understand LLM-generated explanations when the explanations mentioned SQL keywords such as JOIN and GROUP BY or advanced Python programming concepts such as recursion.
    
    \item \textbf{Technical Jargon Clarification:} 
    Participants faced challenges working with technical jargon and concepts central to programming, such as the concepts of ``loop'' and ``recursion'' in Python and the preserved keywords in SQL. Participants sought explanations for technical jargon present either in the source code itself or in the responses of the LLM. 37.50\% of participants in the Text-to-SQL task and 22.22\% in the Python task asked for clarification of technical jargon. Participants often requested clarification after first asking the LLM for a complete explanation. This sequential behavior shows their need for further explanation of the code, suggesting that just asking the LLM for a complete explanations is not sufficient for non-professional programmers to understand the code. Interestingly, in the later phase of the user study (e.g., after the participant completed the first 6-7 questions), a few participants began requesting clarification on technical jargon without asking for complete explanations. This suggests that they were actively learning the code syntax and were able to more directly engage in code debugging as they accumulated more experience. 

    \item \textbf{Execution Process Simulation:}  
    Beyond only requesting explanations from the Chat LLM, participants also asked the LLM for execution process simulation, stepping through how the code actually runs. This method was used in 31.25\% of conversations in the Text-to-SQL task and 38.89\% in the Python task. However, we observed that GPT-3.5 often failed to generate execution steps that fully reflected the logic of the code. When the code included mathematical calculations, the results could be incorrect, even if the execution steps appeared accurate. In some cases, the LLM ``faked'' the final execution result to match the expected output rather than reflecting the true execution. This is an example of a ``LLM Hallucination'', where the LLM generates incorrect or fabricated information. Participants who relied on this method were often misled by the simulation and ultimately failed the coding task. 
    
    \item \textbf{Clarifying Concepts in the Questions:} Participants also sought explanations for unclear concepts within the input questions. This occurred less frequently, 12.5\% in Text-to-SQL and 11.11\% in Python. The low frequency suggests that most participants felt confident in their understanding of the question.
\end{itemize}

\begin{table}[t!]
    \resizebox{0.85\textwidth}{!}{%
    \begin{tabular}{p{0.24\linewidth}|p{0.5\linewidth}|c|c}
    \toprule
    \multirow{2}{*}{\parbox{0.24\textwidth}{\textbf{Methods of Code Comprehension}}} & \multirow{2}{*}{\textbf{Example Participant Input}} & \multicolumn{2}{c}{\textbf{Frequency (\%)}} \\
    \cmidrule{3-4}
    & & Text-to-SQL & Python \\
    \midrule
    \textbf{Complete Explanation} & \textit{"Can you explain what each line of code is doing?"} & 86.75 & 87.22 \\
    \midrule
    \textbf{Technical Jargon Clarification} & \textit{"isinstance(item, list): is that a helper function or an inbuilt function?"} & 37.50              & 22.22 \\
    \midrule
    \textbf{Execution Process Simulation} & \textit{"Show me you doing number 35."} & 31.25 & 38.89 \\
    \midrule
    \textbf{Explanation on Unclear Concept in the Question} & \textit{"Define "predominantly spoken language"."} & 12.50 & 11.11 \\
    \bottomrule     
    \end{tabular} %
    }
    \caption{Various methods of how non-professional programmers interact with a vanilla Chat LLM for code comprehension and its frequency. (Note that the participant may apply multiple methods in one conversation.)}
    \vspace{-\baselineskip}
    \label{tab:gpt_ccd_behavior}
\end{table}

Our analysis reveals that non-professional programmers interacting with a commercial Chat LLM must invent strategies for code comprehension. Due to their lack of programming knowledge, participants struggled to understand the code. However, analysis of participants strategies in understanding code does not directly reveal the ultimate success of these methods in helping participants understand the code. For participants, the key outcome of understanding the code was the ability to successfully ask the LLM to correct the code. Therefore, we next investigated participants' approach to offering corrections to the code.

\subsection{Error Correction} \label{sec:gpt_ec}

\begin{table}[t!]
    \centering
    \resizebox{0.7\textwidth}{!}{%
    \begin{tabular}{l|p{0.1\linewidth}p{0.1\linewidth}p{0.1\linewidth}p{0.1\linewidth}||p{0.18\linewidth}}
    \toprule
    & \multicolumn{4}{p{0.4\linewidth}||}{\textbf{Frequency of Participants Feedback for Error Correction (\%)}} & \textbf{Quality of Participant Feedback (\%)} \\ 
    \cmidrule{2-6} 
    & \multicolumn{1}{c}{\textbf{Easy}} & \multicolumn{1}{c}{\textbf{Medium}} & \multicolumn{1}{c|}{\textbf{Hard}} & \multicolumn{1}{c||}{\textbf{Overall}} & \multicolumn{1}{c}{\textbf{Accuracy}} \\
    \midrule
    \textbf{Text-to-SQL} & \multicolumn{1}{c}{62.50} & \multicolumn{1}{c}{43.75} & \multicolumn{1}{c|}{44.44} & \multicolumn{1}{c||}{51.53} & \multicolumn{1}{c}{38.46} \\
    \textbf{Python Code Gen} & \multicolumn{1}{c}{64.72} & \multicolumn{1}{c}{79.17} & \multicolumn{1}{c|}{53.33} & \multicolumn{1}{c||}{65.64} & \multicolumn{1}{c}{42.22} \\ 
    \bottomrule
    \end{tabular}%
    }
    \caption{Frequency and accuracy of participant feedback for error correction when using the vanilla Chat LLM. Frequency was calculated at the conversation level. Accuracy was calculated based on the first two feedback types listed in Table~\ref{tab:gpt_feedback_sr}, as the other types cannot be precisely measured.
    }
    \vspace{-\baselineskip}
    \label{tab:gpt_ccd_error}
\end{table}

In Table~\ref{tab:gpt_ccd_error}, we present the frequency of participants providing feedback for error correction, as well as the feedback accuracy (defined as the percentage of feedback that precisely instructs the Chat LLM for error correction, which was calculated through manual analysis). As shown in the table, participants were able to propose error correction suggestions in over half of the conversations. At first glance, this might suggest that participants understood the model-generated content sufficiently to identify and address potential errors. However, a closer examination shows that the quality of the participant feedback was low, with only approximately 40\% accuracy for both SQL and Python tasks. The low accuracy of participant feedback highlights substantial gaps in error identification among non-professional programmers when interacting with the vanilla Chat LLM. In Table~\ref{tab:gpt_feedback_sr}, we further summarize the major types of feedback provided by the participants, as well as their frequency and accuracy (except for ``input-output samples'', which are accurate test cases, and ``self-debug'', which refers to participants feeding the same model-predicted code). For each type of feedback, we also present the Chat LLM's SR in error correction \emph{when the feedback is accurate}, so as to ablate the effect of the LLM's capability limit. These feedback types include:

\begin{table}[t!]
\resizebox{0.85\textwidth}{!}{%
    \begin{tabular}{p{0.2\linewidth}|ccc|ccc}
    \toprule
    \multirow{2}{*}{\textbf{Feedback Types}} & \multicolumn{3}{c|}{\textbf{Text-to-SQL}} & \multicolumn{3}{c}{\textbf{Python}} \\
    \cmidrule{2-7} 
    & \textbf{\begin{tabular}[c]{@{}c@{}}Frequency \\ (\%)\end{tabular}} & \textbf{\begin{tabular}[c]{@{}c@{}}Accuracy \\ (\%)\end{tabular}} & \textbf{\begin{tabular}[c]{@{}c@{}}SR for Accurate \\ Feedback (\%)\end{tabular}} & \textbf{\begin{tabular}[c]{@{}c@{}}Frequency \\ (\%)\end{tabular}} & \textbf{\begin{tabular}[c]{@{}c@{}}Accuracy \\ (\%)\end{tabular}} & \textbf{\begin{tabular}[c]{@{}c@{}}SR for Accurate \\ Feedback (\%)\end{tabular}} \\
    \midrule
    \textbf{Instruction for Error Correction} & 71.06 & 42.86 & 60.00  & 67.39 & 42.86 & 94.44 \\ 
    \midrule
    \textbf{Question Rephrasing} & 9.26 & 0.00 & 0.00 & 4.44 & 33.33 & 100.00 \\
    \midrule
    \textbf{Input-Output Samples} & 1.67 & - & 0.00 & 15.06 & - & 12.50 \\ 
    \midrule
    \textbf{Self-Debug} & 25.34 & - & 11.11 & 21.44 & - & 30.00 \\
    \bottomrule
    \end{tabular}%
    }
    \caption{Frequency, accuracy, and success rate (when the feedback is accurate) for different types of participant feedback when interacting with the vanilla Chat LLM. 
    }
    \vspace{-\baselineskip}
    \label{tab:gpt_feedback_sr}
\end{table}

\begin{itemize}
    \item \textbf{Instructions for Error Correction.} The most frequent type of feedback is the instruction pinpointing errors and suggesting fixes to the Chat LLM. For example, the participant may directly point out the LLM's misunderstanding of the question, such as ``official languages are not necessarily predominant.'' However, the low accuracy of these instructions suggests that participants still struggle with code comprehension, making it difficult for them to precisely identify errors. 
    On the other hand, the high SR for accurate feedback indicates that when participants correctly identified issues, their suggestions were effectively taken. This underscores the importance of improving code comprehension to better support non-professional programmers in pinpointing errors. 
    
    \item \textbf{Question Rephrasing.} This type of feedback occurs when participants perceive errors in the generated code and attribute these errors to the underspecified or unclear intent of the original question. Participants attempt to rephrase the question to clarify their intent and guide the LLM towards generating more accurate code. However, due to their limited understanding of programming concepts, many of these rephrasings turned out to be imprecise, which eventually misled the LLM. In the context of Python code generation, only 3 instances of this feedback type were observed, with 1 out of 3 resulting in a correct outcome. This leads to a seemingly higher SR. However, this single instance does not provide enough evidence to confirm the overall effectiveness of question rephrasing as a better feedback strategy than others.
    
    \item \textbf{Input-Output Samples.} In Python code generation, participants also frequently pointed out test examples where the LLM failed and provided the expected output as guidance. However, the absence of detailed instructions to fix the mistakes rendered this type of feedback ineffective (12.5\% SR).
    
    \item \textbf{Self-Debug.} Finally, participants also tried to feed the entire generated code and let the LLM debug the code itself. Same as the input-output samples as feedback, such feedback is not helpful (11\% and 30\% SRs) given that it does not provide any instructive hints on the error correction.
\end{itemize}

Our analysis shows that participants, when interacting with a vanilla Chat LLM, were not able to provide accurate and effective feedback. This inaccuracy and ineffectiveness was caused by both \emph{ineffective code comprehension} and a lack of \emph{structured interaction} between the participant and the Chat LLM. For both code comprehension and error correction, we observed the participants' struggle when there was no structured design to facilitate their interaction with the Chat LLM, and they hence had to invent their strategies, though these strategies were often unhelpful. These findings highlight the need for both \emph{better explanation methods for code comprehension} and \emph{a more structured interaction paradigm} to facilitate non-professional programmers' interactions with the Chat LLM.
\section{\SYSTEM: A Novel Interaction Paradigm with Enhanced NL Explanation} \label{intelliexplain}
In this section, we introduce \SYSTEM, a prompting framework for supporting non-professional programmers in conversational code generation. \SYSTEM features an iterative process where a carefully repurposed Chat LLM explains its generated code, seeks user feedback on the explanation, and refines the code based on that feedback.

\subsection{Design Goals}\label{subsec:design-goals}
Our approach was shaped by two key design goals.

\textbf{(1) Enable non-professional programmers to understand the model-generated code and identify errors \emph{without directly interacting with the code}.} Our study with the vanilla Chat LLM demonstrates that non-professional programmers struggle to understand lengthy and complex code explanations. Sometimes the explanations were also imprecise. These observations suggest the need for explanations that are both accurate and accessible, particularly to allow non-professional programmers to understand the code \emph{without directly reading or interacting with it}. Explanations should also facilitate error identification, enabling users to easily identify defects. 

\textbf{(2) Enable effective incorporation of the non-professional programmer's feedback for error correction.} 
From our user study with the vanilla LLM, we observed that when participants could precisely articulate errors and provide accurate feedback of type ``instruction for error correction,'' the LLM effectively addressed those errors. However, the study also revealed a large variation across different participants' interaction patterns, including the type of feedback they would provide. When the human-LLM interaction is unstructured and fully open-ended, it is not guaranteed that the programmer will always provide the most effective type of feedback. This insight motivated us to design a \emph{structured} interaction paradigm that is well-integrated with our proposed explanation.

In the remaining section, we will first detail our proposed explanation method in Section~\ref{sec:explanation} and then present our designed interaction paradigm in Section~\ref{sec:inter_paradigm}.

\subsection{Enhanced Natural Language Explanations} \label{sec:explanation}

Explanations from the vanilla Chat LLM are often too lengthy and complex to be read by non-professional programmers. To address this limitation, we propose two distinct styles for program explanations for SQL and Python respectively.

\begin{table}[t!]
    \centering
    \resizebox{0.9\columnwidth}{!}{
    \begin{tabular}{p{\columnwidth}}
    \toprule
    Translate the following SQL into question. The question should be consistent with the SQL and follow a similar style as the original question. \\\\

    [...triplets of <SQL, Original Question, Restated Question> as few-shot demonstrations...]\\\\
    
\begin{lstlisting}[language=SQL]
 SELECT status_code FROM bookings GROUP BY status_code ORDER BY count(*) DESC LIMIT 1
\end{lstlisting}

    Original Question: What is the most frequent status of bookings? \\\\
    {Explanation (Restated Question):} Which status code appears most often in bookings?\\
    \bottomrule
    \end{tabular}
    }
    \caption{Restated Question from Source Code as Explanation. The table shows our prompt and an example explanation. Explanation in this format seeks a high-level description of the source code. By comparing the restated question with the original one, users can easily identify any conceptual misunderstanding made by the LLM, which is common in text-to-SQL programming.
    }
    \label{tab:question_restatement}
\end{table}

\subsubsection{Question Restatement from Source Code}
In our user study using the vanilla Chat LLM, we observed that a substantial portion of LLM errors in text-to-SQL tasks originated from the model's misunderstanding of concepts in the user's question. These errors are particularly challenging for non-professional programmers to detect from the code generated, given their limited programming knowledge. The code explanations often contain technical jargon, which can further obscure the underlying concepts and distract users from accurately identifying the LLM's conceptual misunderstanding (Figure~\ref{fig:overview}, right). Observing this challenge, we propose to use a \emph{restated question from the source code} as an explanation for text-to-SQL programming (Table~\ref{tab:question_restatement}). A restated question is an NL question generated by the LLM to describe the intent of a model-generated code. Prior work~\cite{li-etal-2020-mean} found that prompting users to compare the restated question with the original one helps them easily spot any mismatched concepts. For non-professional programmers, these explanations are concise and free of technical jargon. Unlike the previous work which adopted template-based question restatement, we prompt the Chat LLM to generate the question restatement \emph{to follow a similar linguistic pattern as the user's initial question}. Our investigation showed that participants can more easily identify mismatched concepts when the questions to compare follow a similar linguistic structure.

Table~\ref{tab:question_restatement} shows our prompt to the Chat LLM for restated question generation. To align the restated question with the style of the source question, we additionally include the input question in the prompt and explicitly instruct the LLM to produce a restated question following a similar language style. For a more reliable explanation generation, we manually wrote 13 triplets of $<$SQL, Original Question, Restated Question$>$ as few-shot demonstrations to guide the LLM. The SQL queries were selected to encompass a broad range of syntax that may appear in the given programming language, such as keywords ``SELECT'', ``WHERE'', ``DISTINCT'', etc.

\begin{table}[t!]
    \centering
    \resizebox{0.9\columnwidth}{!}{
    \begin{tabular}{p{\columnwidth}}
    \toprule
    You are an expert Python programmer. Your task is to write a description for the following Python program. The description should be accurate, concise, and easily understood by non-programmers.
    \\\\

    [...pairs of <Python Program, Explanation> as few-shot demonstrations...]\\\\
    
    Python Program:
\begin{lstlisting}[language=Python]
import math
def is_not_prime(n):
    result = False
    for i in range(2,int(math.sqrt(n)) + 1):
        if n % i == 0:
            result = True
    return result
\end{lstlisting} \\

    {Explanation (Concise Description):} This program checks if a given number is not a prime number. It does this by iterating through all numbers from 2 to the square root of the given number and checking if any of them divide the number evenly. If a divisor is found, the program returns True, indicating that the number is not prime. Otherwise, it returns False, indicating that the number is prime. \\
    \bottomrule
    \end{tabular}
    }
    \caption{Concise Description of Source Code as Explanation. The table shows our prompt and an example explanation. The explanation includes more details about the thought process behind the source code and thus enables users to identify logic errors in it.
    }
    \label{tab:concise_description}
\end{table}

\subsubsection{Concise Description of Source Code}
In our exploration, we observed that the question restatement as an explanation was more effective for relatively shorter code snippets, such as SQL queries, especially when there is a need to identify conceptual errors in code. However, it does not allow for the detection of logical errors inside the source code, especially in scenarios with lengthy generated code and complex tasks. This issue arises when the LLM makes an inaccurate generation despite correctly understanding concepts in the input question. To address it, we propose a concise explanation of the source code that strikes a balance between the brevity of question restatement and the verbosity of a line-by-line explanation. An example and our prompt design are shown in Table~\ref{tab:concise_description}. We begin by randomly selecting 8 examples and providing each with a human-written description. The description includes a summary of the predicted code at an abstract level, followed by a breakdown of the code logic to illustrate how the LLM approaches the problem. Unlike a line-by-line explanation, our concise description offers a more readable and shorter format while still covering essential logical details. These annotated examples serve as few-shot demonstrations in our prompt to guide the LLM in generating such explanations.

\subsection{\SYSTEM: An Interactive Framework for Conversational Code Generation for Non-Professional Programmers} \label{sec:inter_paradigm}

\begin{figure}[t!]
    \centering
    \includegraphics[width=0.65\linewidth]{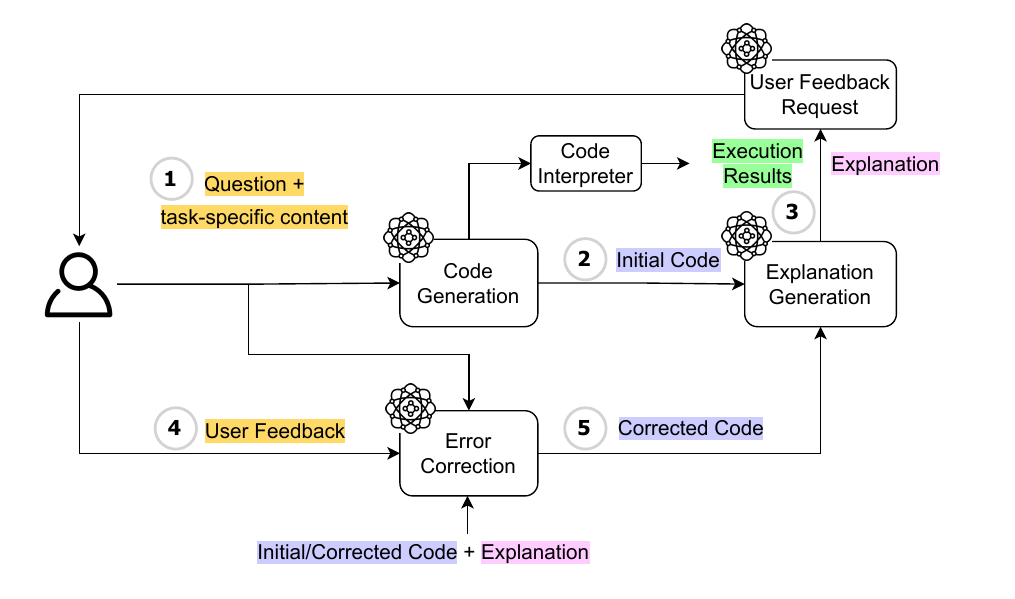}
    \caption{Our proposed interaction paradigm, consisting of (1) a user asks a coding question and provides the context that is necessary for answering the question; (2) LLM predicts an initial code answer; (3) LLM generates an NL explanation for the initial code; (4) the user judges the explanation and determines whether the code is correct; if any error is found in the explanation, the participant provides NL feedback for error correction; and (5) the LLM refines its answer based on the feedback. Steps 3-5 repeat until the user cannot find more errors in the explanation.
    }
    \label{fig:inter_paradigm}
\end{figure}

With our proposed NL explanation, we now present a new interaction paradigm, which is designed to address the variation in how non-professional programmers interact with a Chat LLM and encourage more effective coding task completion (Figure~\ref{fig:inter_paradigm}). This new interaction paradigm leads to \SYSTEM, a new interactive framework that is much more effective in assisting non-professional programmers in conversational code generation tasks. Below, we detail the process of this interaction paradigm.

\vspace{-2mm}
\paragraph{Code Generation and Execution}
We adapt the few-shot prompts from \citet{chen2023teaching} for text-to-SQL and \citet{austin2021program} for Python code generation in \SYSTEM to generate the initial code. Building on previous findings, where a number of participants sought ``execution process simulation'' for code comprehension (Section~\ref{sec:gpt_ccd}), we additionally incorporate an external code interpreter for both tasks. This allows us to execute the generated code to obtain the execution results, which will be utilized by subsequent processes.

\vspace{-2mm}
\paragraph{Explanation Generation}
After the initial code is generated, \SYSTEM prompts the LLM to generate the NL explanation for the code. As introduced in Section~\ref{sec:explanation}, we adopt question restatements as explanations for SQL queries and concise descriptions as explanations for Python code.

\vspace{-2mm}
\paragraph{User Feedback Request}
\SYSTEM then presents the NL explanation and the execution results to the user and seeks feedback on whether their question is correctly answered. In text-to-SQL, participants review a restated question derived from the source code to check if it aligns with their intentions. In Python code generation, users directly assess the correctness of the LLM-generated code by comparing its execution results with the ground truths of the test cases. If the generated code fails any test cases, users identify logical or implementation errors in the code by examining the presented NL explanation. In both tasks, users can mark the question as ``complete'' if no errors are found or provide feedback for error correction.

\vspace{-2mm}
\paragraph{Error Correction}
If feedback for error correction is provided, \SYSTEM refines the code by taking the initial code, its NL explanation, and the user feedback as input. This correction process is formulated using few-shot in-context learning, with 4 human-annotated error correction demonstrations included in the prompt to guide the LLM.

With this interaction paradigm, \SYSTEM iteratively refines the code until participants either identify no further errors or choose to end the conversation.
\section{How Does \SYSTEM Assist Non-Professional Programmers in Coding?}
We conducted an additional user study with 10 participants to evaluate how non-professional programmers interact with \SYSTEM. The recruiting process is the same as those in our study with the vanilla Chat LLM.

\subsection{Overall Performance}
\begin{figure}[t!]
    \centering
    \includegraphics[width=1.0\textwidth]{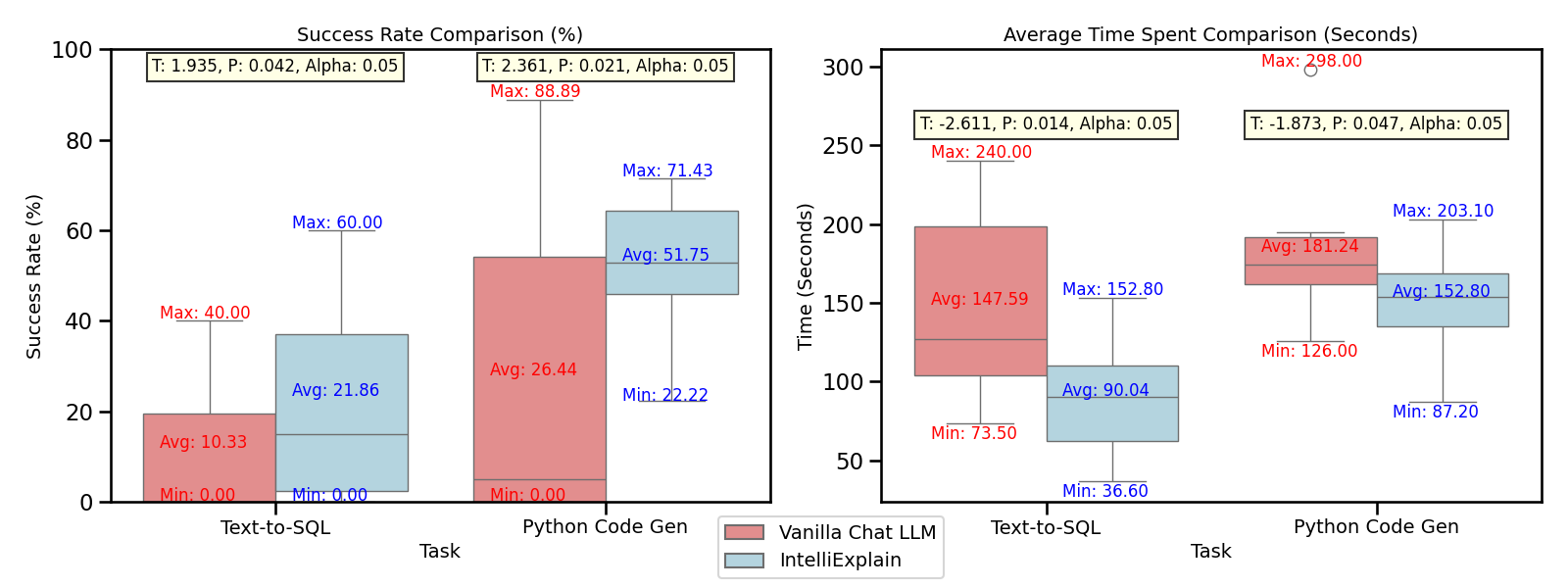}
    \caption{Box plots of the success rate and task time by tool for the Text-to-SQL and Python Code Gen tasks 
    }
    \label{fig:boxplot_t_test}
\end{figure}

Following the same evaluation process of the Chat LLM user study, Table \ref{tab:overall_performance} presents the average success rate (SR) and average time spent per question (Avg. Time/Question) for using \SYSTEM. Similarly, the results excluded skipped samples (6 samples across 4 distinct questions in text-to-SQL and 15 samples across 7 distinct questions in Python) due to ``Cannot understand the question''. \SYSTEM enables participants to achieve SRs 11.53\% and 29.89\%  higher than the vanilla GPT-3.5 group in text-to-SQL and Python code generation, respectively, demonstrating the advantage of our interactive framework. In addition to the improved SR, \SYSTEM also reduced participants' time spent by 60 seconds per question in text-to-SQL and 25 seconds per question in Python code generation. A t-Test (Figure~\ref{fig:boxplot_t_test}) indicated that the difference in means between the two groups are statistically significant, both for SR  (SQL: $t=1.935, p=0.042$; Python: $t=2.361, p=0.021$) and time spent on each question (SQL: $t=-2.611, p=0.014$; Python: $t=-1.873, p=0.047$).

\begin{table}[t!]
\resizebox{0.75\textwidth}{!}{%
    \begin{tabular}{l|cccc}
    \toprule
    \multirow{2}{*}{} & \multicolumn{4}{c}{\textbf{Success Rate (\%)}} \\ \cmidrule{2-5} 
    & \textbf{Easy}  & \textbf{Medium} & \multicolumn{1}{c||}{\textbf{Hard}} & \textbf{Overall} \\ \midrule
    \textbf{Text-to-SQL} & 32.14 {\footnotesize (6.31 $\uparrow$)} & 10.00 {\footnotesize (10.00 $\uparrow$)} & \multicolumn{1}{c||}{17.50 {\footnotesize (17.50 $\uparrow$)}}  & 21.86 {\footnotesize (11.53 $\uparrow$)} ($SD=0.22$) \\
    \textbf{Python Code Gen} & 57.34 {\footnotesize (28.45 $\uparrow$)} & 45.56 {\footnotesize (0.74 $\downarrow$)}  & \multicolumn{1}{c||}{50.48 {\footnotesize (47.15 $\uparrow$)}} & 51.75 {\footnotesize (25.31 $\uparrow$)} ($SD=0.16$) \\ \midrule
    & \multicolumn{4}{c}{\textbf{Ave. Time Spent/Question (s)}} \\ 
    \midrule
    \textbf{Text-to-SQL} & 97.75 {\footnotesize (77.20 $\downarrow$)} & 67.46 {\footnotesize (20.29 $\downarrow$)} & \multicolumn{1}{c||}{93.63 {\footnotesize (56.53 $\downarrow$)}} & 90.04 {\footnotesize (57.55 $\downarrow$)} ($SD=35.97$)  \\
    \textbf{Python Code Gen} & 158.43 {\footnotesize (8.82 $\downarrow$)} & 137.73 {\footnotesize (43.27 $\downarrow$)} & \multicolumn{1}{c||}{160.37 {\footnotesize (39.76 $\downarrow$)}} & 152.80 {\footnotesize (28.44 $\downarrow$)} ($SD=53.09$)  \\
    \bottomrule
    \end{tabular}%
    }
    \caption{Overall participant performance on test questions using \SYSTEM. {We reported the average success rate (in percentage) and time spent (in seconds).} \SYSTEM outperforms vanilla Chat LLM in success rate and generally needs less time ($\uparrow$ denotes an increase compared to the vanilla Chat LLM, while $\downarrow$ represents a decrease).}
    \vspace{-\baselineskip}
    \label{tab:overall_performance}
\end{table}

\begin{figure*}[t!]
    \centering
    \includegraphics[width=0.9\linewidth]{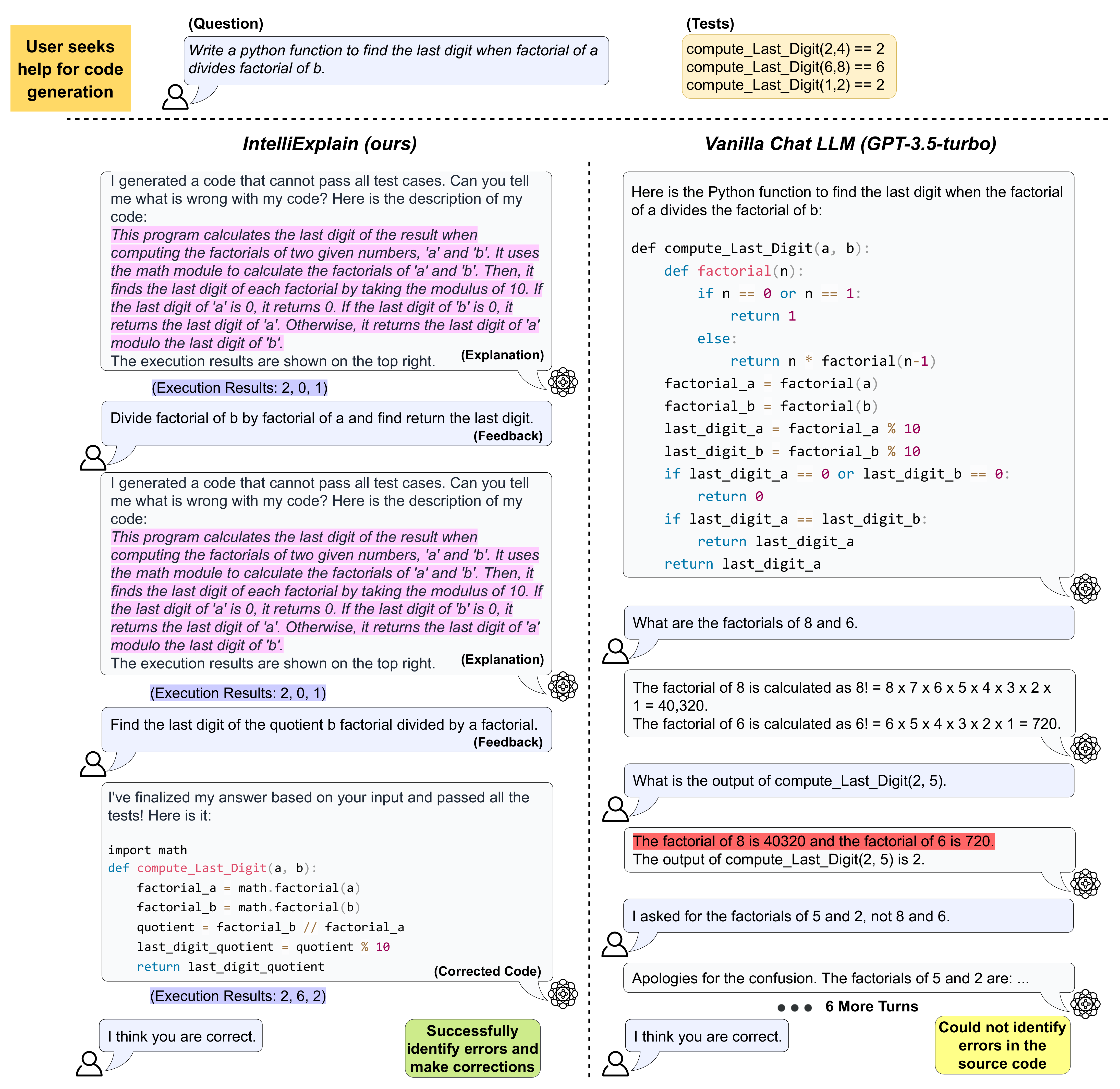}
    \caption{With \SYSTEM, the participant can comprehend the source code via NL explanation (in \textcolor{explainpink}{\faSquare}) to more easily identify potential errors. \SYSTEM makes corrections based on participant feedback. In contrast, when interacting directly with code in vanilla GPT-3.5, the participant struggles to understand source code and fails to identify errors. GPT-3.5 may also sometimes generates responses that are irrelevant to the user question (in \textcolor{explainred}{\faSquare}).}
    \label{fig:case_study_comparsion}
\end{figure*}

We attached one example from our user study for Python code generation in Figure~\ref{fig:case_study_comparsion}; an example of SQL code generation has been presented in Figure~\ref{fig:overview}. In this example, the participant successfully composed the correct Python code using just 2 interactions with \SYSTEM, whereas users relying on the vanilla Chat LLM required significantly more interactions to understand the generated code and validate whether it is correct. Finally, participants using the vanilla Chat LLM failed to identify errors in the generated code, which resulted in an incorrect answer \emph{without the user's awareness}. In the example of \SYSTEM, we noticed that while the participant correctly pointed out the code error in the first turn, their feedback was not incorporated by the framework. \SYSTEM grasped the meaning and made corresponding corrections only when the participant rephrased the feedback in the second turn. While the final outcome is positive, this reveals room for further improvement of \SYSTEM in the future, which we will discuss in Section~\ref{sec:discussion}. In what follows, we look deeply into what exactly makes \SYSTEM work better and quicker than the vanilla Chat LLM.

\subsection{Code Comprehension} \label{sec:ie_ccd}
Through enhanced explanations, \SYSTEM allows participants to understand the code without directly reading or interacting with it, which makes code comprehension much easier for them. To investigate the helpfulness of our enhanced explanations, we manually examined all explanations generated by \SYSTEM, particularly looking at whether they precisely describe the LLM-generated code (Preciseness) and capture code errors (Discriminativeness). As summarized in Table~\ref{tab:explanation_accurate_errors}, our explanations align precisely with the generated code in all cases. We present some of the examples in Table~\ref{tab:explanation_accurate_errors_example}. In most cases (50\% for SQL and 70\% for Python), the explanations are also discriminative enough to allow users to identify code errors. The cases of less or non-discriminative explanations happened more frequently in text-to-SQL tasks, due to the brevity of the restated questions as explanations. While we chose this explanation design to facilitate the identification of code errors caused by conceptual misunderstanding, it compromised the potential of finding other types of code errors from the explanations. In Python code generation, we observed a similar situation. As the code became more complex, concise explanations sometimes failed to capture all the detailed logic and the contained errors. To further examine this problem, we conducted an analysis independent of the user study with a larger sample size. Specifically, we used GPT-3.5 to generate code for the entire Spider-dev set (1,034 test examples) and the MBPP-test set (500 test examples). From the results, we collected 214 erroneous predictions in text-to-SQL and 140 in Python code generation. Then, we prompted GPT-3.5 to generate the explanation for these predictions following our methods in Section~\ref{sec:explanation}. For each task, we randomly selected 30 cases and manually inspected their quality. In text-to-SQL, only one explanation was found to be inconsistent with the source code, while all explanations for the Python code were precise. Among these precise explanations, 51.7\% of the restated questions for text-to-SQL and 66.7\% of the concise explanations for Python code generation successfully captured errors in the source code. These results confirmed our observations from the user study, showing that our explanations are highly precise and largely error-discriminative, although further improvement to balance the conciseness of the explanation with its comprehensiveness is needed.

\begin{table}[t!]
    \resizebox{0.45\textwidth}{!}{%
    \begin{tabular}{l|c|c}
    \toprule
    & \textbf{Preciseness} & \textbf{Discriminativeness}\\
    \midrule
    \textbf{Text-to-SQL} & 10/10 & 5/10\\
    \textbf{Python Code Gen} & 10/10 & 7/10\\
    \bottomrule
    \end{tabular}%
    }
    \caption{The frequency when the NL explanation of \SYSTEM precisely describes the generated code (Preciseness) and when code errors could be found from the explanation (Discriminativeness). Results are calculated based on explanations in the first turn of the participant conversations (hence out of ``10'' explanations).
    }
    \vspace{-\baselineskip}
    \label{tab:explanation_accurate_errors}
\end{table}

\begin{table}[t!]
    \centering\small
    \resizebox{0.95\linewidth}{!}{
    \begin{tabular}{p{0.2\linewidth}p{0.8\linewidth}}
    \toprule
    \textbf{Question (Text-to-SQL)} & How many different winners both participated in the WTA Championships and were left handed? \\\\
    \textbf{Predicted Code} & SELECT COUNT(DISTINCT winner\_id) FROM matches WHERE tourney\_name = 'WTA Championships' AND winner\_hand = 'L' AND \textcolor{red}{loser\_hand != 'L'}\\\\
    \textbf{Explanation} & What is the count of unique winners who were left-handed and participated in the WTA Championships, \textcolor{red}{but their opponents were not left-handed?} \\\\
    \textbf{Errors} & The predicted code include additional condition that is not mentioned in the original question. \\
    \midrule
    \hline
    \textbf{Question (Python Code Gen)} & Write a function to find the kth element in the given array. \\\\
    \textbf{Predicted Code} & \vspace{-3.5mm}
\begin{lstlisting}[language=Python]
def kth_element(arr, n, k):
    arr.sort()
    return arr[k-1]
\end{lstlisting} \\
    \textbf{Explanation} & This program finds the kth smallest element in an array. It takes an array, the size of the array (n), and the value of k as input. \textcolor{red}{The program first sorts the array in ascending order.} Then, it returns the element at the k-1 index position from the sorted array, which represents the kth smallest element.
 \\\\
    \textbf{Errors} & The array does not need to be sorted.\\
    \bottomrule
    \end{tabular}
    }
    \caption{Example explanations that accurately describe the source code and capture \textcolor{red}{errors} existing in the source code.}
    \label{tab:explanation_accurate_errors_example}
    \vspace{-5mm}
\end{table}

\subsection{Error Correction} \label{sec:ie_ec}
\begin{table}[t!]
    \centering
    \resizebox{0.75\textwidth}{!}{%
    \begin{tabular}{l|p{0.1\linewidth}p{0.1\linewidth}p{0.1\linewidth}p{0.1\linewidth}||p{0.18\linewidth}}
    \toprule
    & \multicolumn{4}{p{0.4\linewidth}||}{\textbf{Frequency of Participants Feedback for Error Correction (\%)}} & \textbf{Quality of Participant Feedback (\%)} \\ 
    \cmidrule{2-6} 
    & \multicolumn{1}{c}{\textbf{Easy}} & \multicolumn{1}{c}{\textbf{Medium}} & \multicolumn{1}{c|}{\textbf{Hard}} & \multicolumn{1}{c||}{\textbf{Overall}} & \multicolumn{1}{c}{\textbf{Accuracy}} \\
    \midrule
    \textbf{Text-to-SQL} & \multicolumn{1}{c}{50.71 {\footnotesize (11.79 $\downarrow$)}} & \multicolumn{1}{c}{55.00 {\footnotesize (11.25 $\uparrow$)}} & \multicolumn{1}{c|}{63.61 {\footnotesize (19.17 $\uparrow$)}} & \multicolumn{1}{c||}{56.73 {\footnotesize (5.20 $\uparrow$)}} & \multicolumn{1}{c}{52.83 {\footnotesize (14.37 $\uparrow$)}} \\
    \textbf{Python Code Gen} & \multicolumn{1}{c}{94.44 {\footnotesize (29.72 $\uparrow$)}} & \multicolumn{1}{c}{96.30 {\footnotesize (17.13 $\uparrow$)}} & \multicolumn{1}{c|}{100.00 {\footnotesize (46.67 $\uparrow$)}} & \multicolumn{1}{c||}{96.67 {\footnotesize (31.03 $\uparrow$)}} & \multicolumn{1}{c}{63.86 {\footnotesize (21.64 $\uparrow$)}} \\ 
    \bottomrule
    \end{tabular}%
    }
    \caption{Frequency (at the conversation level) and accuracy of participant feedback for error correction when using \SYSTEM (The $\uparrow$ denotes an increase compared to the vanilla Chat LLM, while the $\downarrow$ represents a decrease).
    }
    \vspace{-\baselineskip}
    \label{tab:ie_error}
\end{table}

We further look into whether our proposed explanations and the structured interaction paradigm make error correction more effective. As illustrated in Table~\ref{tab:ie_error}, participants using \SYSTEM were able to provide corrective feedback based on our enhanced explanations in 56.73\% of text-to-SQL tasks and 96.67\% of Python code generation tasks. This performance surpasses that of the vanilla Chat LLM at nearly all difficulty levels. A natural question here is: did the participants provide more feedback simply because \SYSTEM additionally presented the code execution results (see Section~\ref{sec:inter_paradigm})? In post-task interviews, all 10 participants reported relying on the explanations to correct errors in text-to-SQL tasks, except for one case where a participant noted that the execution results might be incorrect. For Python code generation where test cases and expected outputs were available, 7 out of 10 participants chose to start by reviewing the execution results. They used execution results to decide if the explanations needed careful examination; when the execution results did not meet the expectations, they mainly provided feedback based on the presented explanations. The remaining 3 participants focused on the explanations first, providing feedback immediately when they identified errors, without referring to the execution results. These findings confirmed that, while the additional execution results may make it easier for the participants to identify errors, our enhanced explanations play a critical role in guiding them in providing corrective feedback.

Table~\ref{tab:ie_error} also shows that, in conversations where participants provided feedback for error correction, the accuracy of their feedback increased by 14.37\% and 21.64\% compared to using the vanilla Chat LLM across the two tasks, respectively. These improvements further demonstrate the effectiveness of \SYSTEM in code comprehension and debugging. We list a few examples of participant feedback when using \SYSTEM in Table~\ref{tab:feedback_types}, and present further statistics (i.e., frequency, accuracy, and the success rate for accurate feedback) in Table~\ref{tab:error_correction}. The feedback can be classified into the following three categories:

\begin{table*}[t!]
    \centering
    \resizebox{0.9\linewidth}{!}{
    \begin{tabular}{p{0.25\linewidth}|p{0.75\linewidth}}
    \toprule
    \textbf{Question} & Write a python function to find the last digit when factorial of a divides factorial of b. \\\midrule
    \textbf{Explanation} & This program calculates the last digit of the result when computing the factorials of two given numbers, `a' and `b'. It uses the math module to calculate the factorials of `a' and `b'. \textcolor{red}{Then, it finds the last digit of each factorial by taking the modulus of 10. If the last digit of `a' is 0, it returns 0. If the last digit of `b' is 0, it returns the last digit of `a'. Otherwise, it returns the last digit of `a' modulo the last digit of `b'.} \\
    \bottomrule
    \toprule
    \textbf{Feedback Types} & \textbf{Example} \\
    \midrule
    {\textbf{Instruction for Error Correction}} & After finding the factorials of a and b you should divide the factorial of b by the factorial of a. After dividing you return the last digit from the result.\\
    \midrule
    {\textbf{Question Rephrasing}} & You should write a Python function that determines the last digit of the factorial of number a, which divides the factorial of another number b?\\
    \midrule
    {\textbf{Step-by-Step Instructions}} & First initialize variable fact\_a as the factorial of number `a' and fact\_b as the factorial of number `b'. Then divide fact\_b by fact\_a and assign the value to variable quotient. Then take quotient and return the remainder when divided by 10. \\
    \bottomrule
    \end{tabular}
    }
    \caption{Types of feedback provided by participants. Errors mentioned in the explanation are marked in \textcolor{red}{red}. Diverse types of feedback received from user study demonstrated the effectiveness of our explanation in aiding non-professional programmers in both code comprehension and debugging.}
    \label{tab:feedback_types}
\end{table*}

\begin{itemize}
    \item \textbf{Instructions for Error Correction.} This type of feedback was identified as the most effective in our user study with the vanilla Chat LLM, and it similarly dominated the user study with \SYSTEM. The key difference lies in that participants were able to provide more accurate feedback through our enhanced explanations with improvements of 18.90\% and 29.72\% over the vanilla Chat LLM group.

    \item \textbf{Question Rephrasing.} Participants were more likely to provide this type of feedback in text-to-SQL (39.18\%) compared to Python code generation (2.11\%). This difference is due to the distinct explanation methods used. In text-to-SQL, the restated question encouraged participants to compare the intent of the initial question with the restated one; when inconsistencies were identified, they were more likely to rephrase the question for clarity. In contrast, in Python code generation, participants worked with concise descriptions that included more of the underlying logic, reducing the likelihood of rephrasing the original question. In both tasks, the accuracy of this feedback type is significantly higher (41.18\% for SQL and 16.67\% for Python) than its correspondence in the vanilla Chat LLM's interaction.
    
    \item \textbf{Step-by-Step Instruction.} A new type of feedback was observed in the user study with \SYSTEM appearing in 3.3\% of text-to-SQL and 26.9\% of Python. This feedback involves participants providing detailed, step-by-step instructions to guide the model in solving the problem. Participants, especially those with introductory programming experience, tended to use this feedback when they felt confident in solving the problem themselves. However, the accuracy of this feedback was low, likely due to their limited coding expertise.
    
\end{itemize}

\begin{table}[t!]
\resizebox{0.85\textwidth}{!}{%
    \begin{tabular}{p{0.2\linewidth}|ccc|ccc}
    \toprule
    \multirow{2}{*}{\textbf{Feedback Types}} & \multicolumn{3}{c|}{\textbf{Text-to-SQL}} & \multicolumn{3}{c}{\textbf{Python Code Gen}} \\
    \cmidrule{2-7} 
    & \textbf{\begin{tabular}[c]{@{}c@{}}Frequency \\ (\%)\end{tabular}} & \textbf{\begin{tabular}[c]{@{}c@{}}Accuracy \\ (\%)\end{tabular}} & \textbf{\begin{tabular}[c]{@{}c@{}}SR for Accurate \\ Feedback (\%)\end{tabular}} & \textbf{\begin{tabular}[c]{@{}c@{}}Frequency \\ (\%)\end{tabular}} & \textbf{\begin{tabular}[c]{@{}c@{}}Accuracy \\ (\%)\end{tabular}} & \textbf{\begin{tabular}[c]{@{}c@{}}SR for Accurate \\ Feedback (\%)\end{tabular}} \\
    \midrule
    \textbf{Instruction for Error Correction} & 57.49 & 61.76 {\footnotesize (18.90 $\uparrow$)} & 80.95 {\footnotesize (20.95 $\uparrow$)}  & 71.04 & 74.58 {\footnotesize (31.72 $\uparrow$)} & 88.64 {\footnotesize (5.80 $\downarrow$)} \\ 
    \midrule
    \textbf{Question Rephrasing} & 39.18 & 41.18 & 57.14 & 2.11 & 50.00 & 100.00 \\
    \midrule
    \textbf{Step-by-Step Instructions} & 3.33 & 0.00 & 0.00 & 26.85 & 36.36 & 50.00 \\ 
    \bottomrule
    \end{tabular}
    }
    \caption{Frequency, accuracy, and success rate (when feedback is accurate) of each feedback type when participants interacted with \SYSTEM.
    }
    \vspace{-\baselineskip}
    \label{tab:error_correction}
\end{table}

{The improvement of the participant feedback with \SYSTEM is reflected in not only its standalone accuracy, but also how it makes feedback incorporation easier, as revealed by the better SRs in Table~\ref{tab:error_correction}. Compared to the results with the vanilla Chat LLM (Table~\ref{tab:gpt_feedback_sr}), \SYSTEM's SRs for the overlapped two types of feedback (i.e., ``Instructions for Error Correction'' and ``Question Rephrasing'') are 20.95\% and 57.14\% higher for text-to-SQL task, respectively. For Python code generation, although \SYSTEM's SR for ``Instructions for Error Correction'' was slightly lower than vanilla Chat LLM, this result was based on a significantly larger number of cases (44 conversations with accurate feedback) compared to the vanilla Chat LLM (only 18 conversations with accurate feedback).} We include a more thorough discussion about SRs of ``Instructions for Error Correction'' in Section~\ref{subsec:completeness-of-feedback}. For ``Question Rephrasing'' in text-to-SQL tasks, the SR is still relatively low, due to its limited capacity in pinpointing fine-grained errors (e.g., errors about code logic). In Python code generation, \SYSTEM successfully incorporated ``Step-by-Step Instruction'' feedback, when it was accurate, for 50\% of the cases. {In the remaining unsuccessful cases, the LLM demonstrated difficulties in processing feedback that had unclear logic. Some participants tended to provide step-by-step feedback that was either incomplete or contained too complex logic, making it harder for the system to follow and apply the corrections effectively.}

Overall, the analysis shows that participants were able to provide more effective feedback with the support of our enhanced explanations, as evidenced by significant improvements in feedback accuracy compared to using vanilla GPT-3.5. The absence of uninformative feedback, such as ``input-output sample'' and ``self-debug'' (Section~\ref{sec:gpt_ec}), suggests that our system improves code comprehension and helps users engage more confidently with the LLM. These behaviors were also shaped by our interaction paradigm, which played a crucial role in guiding participants to provide effective feedback. While there are still areas for improvement, particularly in handling more complex codes, these results highlight the potential of well-structured explanations and the interaction paradigm to improve the overall effectiveness of conversational code generation for non-professional programmers.

\subsection{Understanding the Success Rate of ``Instruction for Error Correction'' Feedback} \label{subsec:completeness-of-feedback}

\begin{table}[h!]
    \centering
    \resizebox{0.65\textwidth}{!}{%
    \begin{tabular}{l|p{0.3\textwidth}p{0.3\textwidth}}
        \toprule
        & \multicolumn{2}{p{0.6\textwidth}}{\textbf{Frequency of Complete ``Instruction for Error Correction'' Feedback among Accurate Ones}} \\
        \cmidrule{2-3}
        & \multicolumn{1}{p{0.3\textwidth}|}{\centering \textbf{Text-to-SQL}} & \multicolumn{1}{c}{\textbf{Python}} \\
        \midrule
        \textbf{Vanilla Chat LLM} & \multicolumn{1}{c|}{60.00\%}  & \multicolumn{1}{c}{27.78\%} \\
        \midrule
        \textbf{\SYSTEM} & \multicolumn{1}{c|}{71.43\%} & \multicolumn{1}{c}{72.73\%} \\ 
        \bottomrule
    \end{tabular}%
    }
    \caption{Frequency of complete feedback instances out of the accurate ones for the type of ``Instruction for Error Correction''. Numbers were measured by counting the number of feedback instances that addressed all errors in the code out of all accurate feedback provided.}
    \vspace{-\baselineskip}
    \label{tab:completeness_feedback}
\end{table}

\begin{table}[t!]
    \centering
    \resizebox{0.65\textwidth}{!}{%
    \begin{tabular}{l|p{0.3\textwidth}p{0.3\textwidth}}
        \toprule
        & \multicolumn{2}{p{0.6\textwidth}}{\textbf{
        Percentage of Success Cases with Complete Feedback for ``Instruction for Error Correction'' Type}
        } \\
        \cmidrule{2-3}
        & \multicolumn{1}{p{0.3\textwidth}|}{\centering \textbf{Text-to-SQL}} & \multicolumn{1}{c}{\textbf{Python Code Gen}} \\
        \midrule
        \textbf{Percentage} & \multicolumn{1}{c|}{82.35\%}  & \multicolumn{1}{c}{74.36\%} \\
        \bottomrule
    \end{tabular}%
    }
    \caption{ 
    Percentage of successful error corrections with \SYSTEM that were based on complete feedback, calculated for the feedback type of ``Instruction for Error Correction''. Note that all successful cases in our study had at least accurate feedback.}
    \vspace{-\baselineskip}
    \label{tab:completeness_feedback_sr}
\end{table}

To gain a deeper understanding of the success rate of accurate feedback with \SYSTEM, we conducted a follow-up analysis focusing on the ``Instruction for Error Correction'' type, as it was the most frequent type in both tasks. Specifically, for accurate feedback, we further assessed its \emph{completeness}, i.e., whether the feedback comprehensively addressed all errors present in the code. We calculate the frequency of complete feedback among accurate ones. The results in Table~\ref{tab:completeness_feedback} show that \SYSTEM significantly outperforms the vanilla Chat LLM. This highlights the critical role of our enhanced explanations in assisting non-professional programmers in providing high-quality (accurate and complete) feedback.

In Table~\ref{tab:completeness_feedback_sr}, we further calculate, among the successful error correction cases with \SYSTEM, what percentage of them were based on complete and accurate feedback. We note that all success cases came with at least accurate feedback. The percentages shown in the table thus indicate to what extent the complete feedback is necessary for successful error correction using \SYSTEM.
For both tasks, we observed a clear relationship between the completeness of user feedback and the success rate of error correction using \SYSTEM. 
82.35\% and 74.36\% successful corrections occurred when the feedback was both accurate and complete in SQL and Python tasks, respectively. This observation between feedback completeness and success rate highlights the importance of providing thorough and detailed feedback to the LLM for effective error correction. On the other hand, an analysis of the remaining successful cases with incomplete feedback revealed an interesting finding: in situations where the generated code required a complete rewrite, even when participants provided only partial or incomplete feedback, the LLM was sometimes able to produce correct code during the reconstruction process.

Despite these improvements, there is still room for enhancing the completeness of feedback. Several factors may contribute to this gap. First, the higher complexity of codes might have led to misunderstandings or incomplete explanations, which in turn affected both the accuracy and the completeness of the feedback. Second, a participant's prior knowledge and experience influenced their ability to interpret an explanation and provide helpful feedback. Future research could focus on creating more intuitive human-system interaction designs that help users better understand complex code and provide more complete feedback.

In addition to examining successful cases, we also investigated instances where participants provided accurate and complete feedback for error correction, but the LLM failed to implement the corrections into the code. We identified 4 such cases in the user study of \SYSTEM across two tasks. Although the number of failures is relatively small, this finding underscores a potential limitation of the current Chat LLM, namely its inability to consistently follow human instructions, even when those instructions are clear and precise. A potential reason is that the LLM might struggle with understanding the intent behind the instructions, particularly if the language used by the participant is slightly ambiguous or deviates from the language patterns that the LLM was pre-trained on. This sensitivity highlights the need for further improvement of LLMs to better interpret and follow human instructions across a variety of formats and linguistic styles. 
\section{Performance of \SYSTEM with GPT-4 as Backbone LLM}

\begin{table}[h]
    \centering\small
    \resizebox{0.55\textwidth}{!}{
    \begin{tabular}{l|cc}
    \toprule
    \multirow{2}{*}{} & \multicolumn{2}{c}{\textbf{Performance of \SYSTEM (GPT-4)}} \\
    \cmidrule{2-3}
    & \textbf{Success Rate (\%)} & \textbf{Avg. Time/question (s)} \\
    \midrule
    \textbf{Text-to-SQL} & 30.00 & 107.37 \\
    \textbf{Python Code Gen} & 50.00 & 132.25 \\
    \bottomrule
    \end{tabular}
    }
    \caption{Overall performance of \SYSTEM\ with GPT-4 as backbone Chat LLM. The result was collected based on a pilot study with one non-professional programmer participant.}
    \vspace{-\baselineskip}
    \label{tab:gpt4_result}
\end{table}

\begin{table*}[t!]
    \centering
    \resizebox{0.9\linewidth}{!}{
    \begin{tabular}{p{0.25\linewidth}p{0.75\linewidth}}
    \toprule
    \textbf{Question} & Write a function to find the depth of a dictionary. \\
    \vspace{2.5mm}\textbf{Predited Code} & 
\begin{lstlisting}[language=Python]
def dict_depth(dictionary):
    depth = 0
    if not isinstance(dictionary, dict) or len(dictionary) == 0:
        return depth
    else:
        depth += 1
        return depth + max(dict_depth(value) for value in dictionary.values())
\end{lstlisting}\\
    \midrule
    \textbf{GPT-3.5 Explanation} & This program calculates the depth or level of nesting in a given dictionary. It checks if the input is a dictionary and if it is not empty. If it is not a dictionary or is empty, it returns a depth of 0. Otherwise, it recursively calculates the depth of each value in the dictionary and returns the maximum depth found, incremented by 1.\\
    \midrule
    \textbf{GPT-4 Explanation} & This program calculates the depth of a dictionary. \highlight{98D418}{The depth of a dictionary is the maximum number of levels it contains. If the dictionary contains other dictionaries as values, those are considered additional levels.} The program checks each value in the dictionary, and if a value is another dictionary, it recursively calculates the depth of that dictionary. The final depth is one more than the maximum depth of any value. If the dictionary is empty, its depth is considered to be zero.\\
    \bottomrule
    \end{tabular}
    }
    \caption{Explanations provided by GPT-3.5 and GPT-4 on the same question for the generated code. GPT-4 generates an explanatory sentence on how to solve the problem in general (in \textcolor{explaingreen}{\faSquare}). This information makes the explanation generated by GPT-4 more comprehensible than GPT-3.5.
    }
    \label{tab:explanation_gpt3_gpt4}
\end{table*}

Our main investigation has been based on GPT-3.5-turbo rather than the state-of-the-art GPT-4 model. A natural question here is thus: for the effectiveness we have shown with \SYSTEM in our study, as well as the findings we have discovered, are they still applicable when people switch to the more powerful Chat LLM backend? Limited by the available resources and budget, it is infeasible to re-conduct the user study. However, to gain some preliminary insights, we still performed a pilot study with one participant who had no prior experience in SQL and Python programming. Given GPT-4's enhanced code generation capabilities, some questions we used in the main user study were not suitable anymore. Specifically, we observed that 3 out of 10 questions in text-to-SQL and 2 out of 10 questions in Python code generation could be accurately answered by GPT-4 without any specific interaction design. Consequently, we excluded these questions and randomly selected additional questions to keep the same amount of questions. The results are presented in Table~\ref{tab:gpt4_result}. The results show an improved success rate in text-to-SQL and a comparable success rate in Python code generation compared to \SYSTEM (GPT-3.5)'s (Table~\ref{tab:overall_performance}), which indicates that a stronger Chat LLM could potentially yield even more effective human-LLM interaction for code generation. The results also demonstrate that our designed prompts and interaction paradigm can work with a more powerful Chat LLM.

To gain deeper insights into the differences between GPT-4 and GPT-3.5, we conducted an analysis focusing on the quality of their generated NL explanations. As expected, we observed that GPT-4 could produce explanations as precise as GPT-3.5's. Beyond preciseness, we found that in two cases, the explanations generated by GPT-4 show even higher quality in terms of comprehensibility, as exemplified in Table~\ref{tab:explanation_gpt3_gpt4}. This enhanced comprehensibility could potentially improve the user experience with clearer insights into the generated code. However, the observation that only two explanations reveal this enhanced comprehensibility indicates that there is still room for improvement even with a more advanced Chat LLM. We have also examined whether the participant could provide effective feedback for error correction. In the pilot study, the participant mainly provided feedback of type ``Instruction for Error Correction'', except one ``Question Rephrasing'' feedback for text-to-SQL. The observation reaffirms that with our designed explanations, users can find errors and provide feedback without directly interacting with the source code. However, its low success rate underscores the need for future exploration on this topic. 
\section{Discussion} \label{sec:discussion}
Our user study with the vanilla Chat LLM revealed key insights into how non-professional programmers interact with LLMs for conversational code generation. Building on these findings, we proposed \SYSTEM, which introduced an improved interaction paradigm with enhanced explanations and a more effective feedback loop. While these enhancements have shown promising results, it's important to acknowledge the limitations in success rates and explore potential future improvements that could further advance this framework.

\textbf{Improving Explanation to Handle More Complex Code Logic:} The success rate in both user studies highly depends on the accuracy and completeness of human feedback. Our current design of code explanations shows limitations in fully describing the underlying logic of more complex code. These shortcomings can hinder users' ability to provide precise feedback, which is crucial for effective error correction. Although LLM-generated step-by-step explanations might appear comprehensive, they often fall short, as their excessive length and complexity can confuse non-professional programmers. To address these issues, new design considerations for explanations should focus on clarity and conciseness. Explanations need to strike a balance between being detailed enough to capture the code's logic and simple enough to be easily understood by users without sufficient expertise. This approach should reduce cognitive load, enabling users to engage more effectively with the explanations and provide more accurate feedback. Additionally, the design should include mechanisms to highlight the most critical aspects of the code, ensuring that users can quickly grasp the essential elements without getting lost in unnecessary details.

\textbf{Interactable Explanations and Feedback:}
From post-task interviews using \SYSTEM, we noticed that some participants could not notice small changes in the explanations, which led to incorrect error corrections. The differences between code iterations and their corresponding explanations could be more explicit. This could include visually highlighting changes between turns, so users can easily track adjustments and understand the changes in the code. Additionally, an actionable mechanism that shows how different parts of user feedback impact code generation would be beneficial. This  would allow users to see the immediate effects of their input, providing a clearer understanding of its effectiveness and helping them refine their feedback for better results.

\textbf{Implementing Mechanism of LLMs Seeking Clarifications on Uncertain Concepts in User Input:}
{In addition to analyzing how user experience in code comprehension and error correction can be improved in conversational code generation, we also examined why the Chat LLM fail to generate correct code following human instructions. Our analysis of unsuccessful conversations both with and without user feedback for error correction showed that the Chat LLM struggled with ambiguous user input, where the description of the input question or user feedback was unclear in some extent. Rather than seeking clarification on unclear concepts, the Chat LLM often proceeded based on incorrect assumptions, which led to incorrect responses. This observation suggests a need for mechanisms that allow the LLM to recognize confusion and request clarification when it encounters ambiguous input.}

\textbf{LLM Hallucination:} As discussed in Section~\ref{sec:gpt_ccd}, we observed hallucinated behaviors in the vanilla Chat LLM, such as generating code execution processes that are not logically aligned with the code or providing ``fake'' execution results. In Section~\ref{sec:ie_ccd}, we systematically examined \SYSTEM for the quality of its generated explanations and found no hallucination issues. We attributed this success to our careful prompt design in Section~\ref{sec:explanation}. However, we note that slight hallucination was still found in other parts of \SYSTEM's function; for example, we observed one SQL query where \SYSTEM did not join two tables based on the correct keys.
\section{Conclusion}
In this work, we systematically explored the usability and limitations of Chat LLMs in helping non-professional programmers with conversational code generation. We identified challenges in the vanilla Chat LLM and proposed a new structured interaction paradigm with enhanced explanations to address these issues. Our results show that the improved explanations help users better understand and debug code, allowing them to provide more accurate feedback. The interactive feedback loop also effectively refines the code based on user input, leading to higher success rates and a better overall experience compared to the vanilla LLM.
\begin{acks}
This project was sponsored by NSF SHF 2311468, GMU College of Computing and Engineering, and GMU Department of Computer Science. We appreciate the Office of Research Integrity and Assurance at GMU for their work in reviewing and approving our Institutional Review Board (IRB) application. We also appreciate comments from students in GMU NLP and SE labs.
\end{acks}

\newpage
\bibliographystyle{ACM-Reference-Format}
\bibliography{custom}


\begin{thebibliography}{52}


\ifx \showCODEN    \undefined \def \showCODEN     #1{\unskip}     \fi
\ifx \showDOI      \undefined \def \showDOI       #1{#1}\fi
\ifx \showISBNx    \undefined \def \showISBNx     #1{\unskip}     \fi
\ifx \showISBNxiii \undefined \def \showISBNxiii  #1{\unskip}     \fi
\ifx \showISSN     \undefined \def \showISSN      #1{\unskip}     \fi
\ifx \showLCCN     \undefined \def \showLCCN      #1{\unskip}     \fi
\ifx \shownote     \undefined \def \shownote      #1{#1}          \fi
\ifx \showarticletitle \undefined \def \showarticletitle #1{#1}   \fi
\ifx \showURL      \undefined \def \showURL       {\relax}        \fi
\providecommand\bibfield[2]{#2}
\providecommand\bibinfo[2]{#2}
\providecommand\natexlab[1]{#1}
\providecommand\showeprint[2][]{arXiv:#2}

\bibitem[Achiam et~al\mbox{.}(2023)]%
        {achiam2023gpt}
\bibfield{author}{\bibinfo{person}{Josh Achiam}, \bibinfo{person}{Steven Adler}, \bibinfo{person}{Sandhini Agarwal}, \bibinfo{person}{Lama Ahmad}, \bibinfo{person}{Ilge Akkaya}, \bibinfo{person}{Florencia~Leoni Aleman}, \bibinfo{person}{Diogo Almeida}, \bibinfo{person}{Janko Altenschmidt}, \bibinfo{person}{Sam Altman}, \bibinfo{person}{Shyamal Anadkat}, {et~al\mbox{.}}} \bibinfo{year}{2023}\natexlab{}.
\newblock \showarticletitle{Gpt-4 technical report}.
\newblock \bibinfo{journal}{\emph{arXiv preprint arXiv:2303.08774}} (\bibinfo{year}{2023}).
\newblock


\bibitem[AI(2024)]%
        {gemini}
\bibfield{author}{\bibinfo{person}{Google AI}.} \bibinfo{year}{2024}\natexlab{}.
\newblock \bibinfo{title}{Gemini}.
\newblock
\newblock
\urldef\tempurl%
\url{https://www.example.com/gemini}
\showURL{%
\tempurl}
\newblock
\shownote{Large language model}.


\bibitem[{Anthropic}(2024)]%
        {claude}
\bibfield{author}{\bibinfo{person}{{Anthropic}}.} \bibinfo{year}{2024}\natexlab{}.
\newblock \bibinfo{title}{Claude}.
\newblock \bibinfo{howpublished}{\url{https://www.anthropic.com}}.
\newblock
\newblock
\shownote{Large language model}.


\bibitem[Austin et~al\mbox{.}(2021)]%
        {austin2021program}
\bibfield{author}{\bibinfo{person}{Jacob Austin}, \bibinfo{person}{Augustus Odena}, \bibinfo{person}{Maxwell Nye}, \bibinfo{person}{Maarten Bosma}, \bibinfo{person}{Henryk Michalewski}, \bibinfo{person}{David Dohan}, \bibinfo{person}{Ellen Jiang}, \bibinfo{person}{Carrie Cai}, \bibinfo{person}{Michael Terry}, \bibinfo{person}{Quoc Le}, {et~al\mbox{.}}} \bibinfo{year}{2021}\natexlab{}.
\newblock \showarticletitle{Program synthesis with large language models}.
\newblock \bibinfo{journal}{\emph{arXiv preprint arXiv:2108.07732}} (\bibinfo{year}{2021}).
\newblock


\bibitem[Barke et~al\mbox{.}(2023)]%
        {barke2023grounded}
\bibfield{author}{\bibinfo{person}{Shraddha Barke}, \bibinfo{person}{Michael~B James}, {and} \bibinfo{person}{Nadia Polikarpova}.} \bibinfo{year}{2023}\natexlab{}.
\newblock \showarticletitle{Grounded copilot: How programmers interact with code-generating models}.
\newblock \bibinfo{journal}{\emph{Proceedings of the ACM on Programming Languages}} \bibinfo{volume}{7}, \bibinfo{number}{OOPSLA1} (\bibinfo{year}{2023}), \bibinfo{pages}{85--111}.
\newblock


\bibitem[Becker et~al\mbox{.}(2023)]%
        {becker2023programming}
\bibfield{author}{\bibinfo{person}{Brett~A Becker}, \bibinfo{person}{Paul Denny}, \bibinfo{person}{James Finnie-Ansley}, \bibinfo{person}{Andrew Luxton-Reilly}, \bibinfo{person}{James Prather}, {and} \bibinfo{person}{Eddie~Antonio Santos}.} \bibinfo{year}{2023}\natexlab{}.
\newblock \showarticletitle{Programming is hard-or at least it used to be: Educational opportunities and challenges of ai code generation}. In \bibinfo{booktitle}{\emph{Proceedings of the 54th ACM Technical Symposium on Computer Science Education V. 1}}. \bibinfo{pages}{500--506}.
\newblock


\bibitem[Champa et~al\mbox{.}(2024)]%
        {champa2024chatgpt}
\bibfield{author}{\bibinfo{person}{Arifa~Islam Champa}, \bibinfo{person}{Md~Fazle Rabbi}, \bibinfo{person}{Costain Nachuma}, {and} \bibinfo{person}{Minhaz~F Zibran}.} \bibinfo{year}{2024}\natexlab{}.
\newblock \showarticletitle{ChatGPT in action: Analyzing its use in software development}. In \bibinfo{booktitle}{\emph{Proceedings of the 21st International Conference on Mining Software Repositories}}. \bibinfo{pages}{182--186}.
\newblock


\bibitem[Chaurasia and Mooney(2017)]%
        {chaurasia-mooney-2017-dialog}
\bibfield{author}{\bibinfo{person}{Shobhit Chaurasia} {and} \bibinfo{person}{Raymond~J. Mooney}.} \bibinfo{year}{2017}\natexlab{}.
\newblock \showarticletitle{Dialog for Language to Code}. In \bibinfo{booktitle}{\emph{Proceedings of the Eighth International Joint Conference on Natural Language Processing (Volume 2: Short Papers)}}, \bibfield{editor}{\bibinfo{person}{Greg Kondrak} {and} \bibinfo{person}{Taro Watanabe}} (Eds.). \bibinfo{publisher}{Asian Federation of Natural Language Processing}, \bibinfo{address}{Taipei, Taiwan}, \bibinfo{pages}{175--180}.
\newblock
\urldef\tempurl%
\url{https://aclanthology.org/I17-2030}
\showURL{%
\tempurl}


\bibitem[Chen et~al\mbox{.}(2023b)]%
        {chen2023improving}
\bibfield{author}{\bibinfo{person}{Angelica Chen}, \bibinfo{person}{J{\'e}r{\'e}my Scheurer}, \bibinfo{person}{Tomasz Korbak}, \bibinfo{person}{Jon~Ander Campos}, \bibinfo{person}{Jun~Shern Chan}, \bibinfo{person}{Samuel~R Bowman}, \bibinfo{person}{Kyunghyun Cho}, {and} \bibinfo{person}{Ethan Perez}.} \bibinfo{year}{2023}\natexlab{b}.
\newblock \showarticletitle{Improving code generation by training with natural language feedback}.
\newblock \bibinfo{journal}{\emph{arXiv preprint arXiv:2303.16749}} (\bibinfo{year}{2023}).
\newblock


\bibitem[Chen et~al\mbox{.}(2021)]%
        {chen2021evaluating}
\bibfield{author}{\bibinfo{person}{Mark Chen}, \bibinfo{person}{Jerry Tworek}, \bibinfo{person}{Heewoo Jun}, \bibinfo{person}{Qiming Yuan}, \bibinfo{person}{Henrique Ponde de~Oliveira Pinto}, \bibinfo{person}{Jared Kaplan}, \bibinfo{person}{Harri Edwards}, \bibinfo{person}{Yuri Burda}, \bibinfo{person}{Nicholas Joseph}, \bibinfo{person}{Greg Brockman}, {et~al\mbox{.}}} \bibinfo{year}{2021}\natexlab{}.
\newblock \showarticletitle{Evaluating large language models trained on code}.
\newblock \bibinfo{journal}{\emph{arXiv preprint arXiv:2107.03374}} (\bibinfo{year}{2021}).
\newblock


\bibitem[Chen et~al\mbox{.}(2023a)]%
        {chen2023teaching}
\bibfield{author}{\bibinfo{person}{Xinyun Chen}, \bibinfo{person}{Maxwell Lin}, \bibinfo{person}{Nathanael Sch{\"a}rli}, {and} \bibinfo{person}{Denny Zhou}.} \bibinfo{year}{2023}\natexlab{a}.
\newblock \showarticletitle{Teaching large language models to self-debug}.
\newblock \bibinfo{journal}{\emph{arXiv preprint arXiv:2304.05128}} (\bibinfo{year}{2023}).
\newblock


\bibitem[Chopra et~al\mbox{.}(2023)]%
        {chopra2023conversational}
\bibfield{author}{\bibinfo{person}{Bhavya Chopra}, \bibinfo{person}{Ananya Singha}, \bibinfo{person}{Anna Fariha}, \bibinfo{person}{Sumit Gulwani}, \bibinfo{person}{Chris Parnin}, \bibinfo{person}{Ashish Tiwari}, {and} \bibinfo{person}{Austin~Z Henley}.} \bibinfo{year}{2023}\natexlab{}.
\newblock \showarticletitle{Conversational challenges in ai-powered data science: Obstacles, needs, and design opportunities}.
\newblock \bibinfo{journal}{\emph{arXiv preprint arXiv:2310.16164}} (\bibinfo{year}{2023}).
\newblock


\bibitem[Drosos et~al\mbox{.}(2020)]%
        {drosos2020wrex}
\bibfield{author}{\bibinfo{person}{Ian Drosos}, \bibinfo{person}{Titus Barik}, \bibinfo{person}{Philip~J Guo}, \bibinfo{person}{Robert DeLine}, {and} \bibinfo{person}{Sumit Gulwani}.} \bibinfo{year}{2020}\natexlab{}.
\newblock \showarticletitle{Wrex: A unified programming-by-example interaction for synthesizing readable code for data scientists}. In \bibinfo{booktitle}{\emph{Proceedings of the 2020 CHI conference on human factors in computing systems}}. \bibinfo{pages}{1--12}.
\newblock


\bibitem[Elgohary et~al\mbox{.}(2020)]%
        {elgohary-etal-2020-speak}
\bibfield{author}{\bibinfo{person}{Ahmed Elgohary}, \bibinfo{person}{Saghar Hosseini}, {and} \bibinfo{person}{Ahmed Hassan~Awadallah}.} \bibinfo{year}{2020}\natexlab{}.
\newblock \showarticletitle{Speak to your Parser: Interactive Text-to-{SQL} with Natural Language Feedback}. In \bibinfo{booktitle}{\emph{Proceedings of the 58th Annual Meeting of the Association for Computational Linguistics}}, \bibfield{editor}{\bibinfo{person}{Dan Jurafsky}, \bibinfo{person}{Joyce Chai}, \bibinfo{person}{Natalie Schluter}, {and} \bibinfo{person}{Joel Tetreault}} (Eds.). \bibinfo{publisher}{Association for Computational Linguistics}, \bibinfo{address}{Online}, \bibinfo{pages}{2065--2077}.
\newblock
\urldef\tempurl%
\url{https://doi.org/10.18653/v1/2020.acl-main.187}
\showDOI{\tempurl}


\bibitem[Elgohary et~al\mbox{.}(2021)]%
        {elgohary-etal-2021-nl}
\bibfield{author}{\bibinfo{person}{Ahmed Elgohary}, \bibinfo{person}{Christopher Meek}, \bibinfo{person}{Matthew Richardson}, \bibinfo{person}{Adam Fourney}, \bibinfo{person}{Gonzalo Ramos}, {and} \bibinfo{person}{Ahmed~Hassan Awadallah}.} \bibinfo{year}{2021}\natexlab{}.
\newblock \showarticletitle{{NL}-{EDIT}: Correcting Semantic Parse Errors through Natural Language Interaction}. In \bibinfo{booktitle}{\emph{Proceedings of the 2021 Conference of the North American Chapter of the Association for Computational Linguistics: Human Language Technologies}}, \bibfield{editor}{\bibinfo{person}{Kristina Toutanova}, \bibinfo{person}{Anna Rumshisky}, \bibinfo{person}{Luke Zettlemoyer}, \bibinfo{person}{Dilek Hakkani-Tur}, \bibinfo{person}{Iz~Beltagy}, \bibinfo{person}{Steven Bethard}, \bibinfo{person}{Ryan Cotterell}, \bibinfo{person}{Tanmoy Chakraborty}, {and} \bibinfo{person}{Yichao Zhou}} (Eds.). \bibinfo{publisher}{Association for Computational Linguistics}, \bibinfo{address}{Online}, \bibinfo{pages}{5599--5610}.
\newblock
\urldef\tempurl%
\url{https://doi.org/10.18653/v1/2021.naacl-main.444}
\showDOI{\tempurl}


\bibitem[Ge and Wu(2023)]%
        {ge2023empirical}
\bibfield{author}{\bibinfo{person}{Haotong Ge} {and} \bibinfo{person}{Yuemeng Wu}.} \bibinfo{year}{2023}\natexlab{}.
\newblock \showarticletitle{An Empirical Study of Adoption of ChatGPT for Bug Fixing among Professional Developers}.
\newblock \bibinfo{journal}{\emph{Innovation \& Technology Advances}} \bibinfo{volume}{1}, \bibinfo{number}{1} (\bibinfo{year}{2023}), \bibinfo{pages}{21--29}.
\newblock


\bibitem[GitHub(2021)]%
        {github:copilot}
\bibfield{author}{\bibinfo{person}{GitHub}.} \bibinfo{year}{2021}\natexlab{}.
\newblock \bibinfo{title}{Copilot}.
\newblock \bibinfo{howpublished}{\url{https://github.blog/2021-06-29-introducing-github-copilot-ai-pair-programmer/}}.
\newblock


\bibitem[Gur et~al\mbox{.}(2018)]%
        {gur-etal-2018-dialsql}
\bibfield{author}{\bibinfo{person}{Izzeddin Gur}, \bibinfo{person}{Semih Yavuz}, \bibinfo{person}{Yu Su}, {and} \bibinfo{person}{Xifeng Yan}.} \bibinfo{year}{2018}\natexlab{}.
\newblock \showarticletitle{{D}ial{SQL}: Dialogue Based Structured Query Generation}. In \bibinfo{booktitle}{\emph{Proceedings of the 56th Annual Meeting of the Association for Computational Linguistics (Volume 1: Long Papers)}}, \bibfield{editor}{\bibinfo{person}{Iryna Gurevych} {and} \bibinfo{person}{Yusuke Miyao}} (Eds.). \bibinfo{publisher}{Association for Computational Linguistics}, \bibinfo{address}{Melbourne, Australia}, \bibinfo{pages}{1339--1349}.
\newblock
\urldef\tempurl%
\url{https://doi.org/10.18653/v1/P18-1124}
\showDOI{\tempurl}


\bibitem[Kazemitabaar et~al\mbox{.}(2023a)]%
        {kazemitabaar2023studying}
\bibfield{author}{\bibinfo{person}{Majeed Kazemitabaar}, \bibinfo{person}{Justin Chow}, \bibinfo{person}{Carl Ka~To Ma}, \bibinfo{person}{Barbara~J Ericson}, \bibinfo{person}{David Weintrop}, {and} \bibinfo{person}{Tovi Grossman}.} \bibinfo{year}{2023}\natexlab{a}.
\newblock \showarticletitle{Studying the effect of AI Code Generators on Supporting Novice Learners in Introductory Programming}. In \bibinfo{booktitle}{\emph{Proceedings of the 2023 CHI Conference on Human Factors in Computing Systems}}. \bibinfo{pages}{1--23}.
\newblock


\bibitem[Kazemitabaar et~al\mbox{.}(2023b)]%
        {kazemitabaar2023novices}
\bibfield{author}{\bibinfo{person}{Majeed Kazemitabaar}, \bibinfo{person}{Xinying Hou}, \bibinfo{person}{Austin Henley}, \bibinfo{person}{Barbara~Jane Ericson}, \bibinfo{person}{David Weintrop}, {and} \bibinfo{person}{Tovi Grossman}.} \bibinfo{year}{2023}\natexlab{b}.
\newblock \showarticletitle{How novices use LLM-based code generators to solve CS1 coding tasks in a self-paced learning environment}. In \bibinfo{booktitle}{\emph{Proceedings of the 23rd Koli Calling International Conference on Computing Education Research}}. \bibinfo{pages}{1--12}.
\newblock


\bibitem[Khojah et~al\mbox{.}(2024)]%
        {khojah2024beyond}
\bibfield{author}{\bibinfo{person}{Ranim Khojah}, \bibinfo{person}{Mazen Mohamad}, \bibinfo{person}{Philipp Leitner}, {and} \bibinfo{person}{Francisco~Gomes de Oliveira~Neto}.} \bibinfo{year}{2024}\natexlab{}.
\newblock \showarticletitle{Beyond code generation: An observational study of chatgpt usage in software engineering practice}.
\newblock \bibinfo{journal}{\emph{Proceedings of the ACM on Software Engineering}} \bibinfo{volume}{1}, \bibinfo{number}{FSE} (\bibinfo{year}{2024}), \bibinfo{pages}{1819--1840}.
\newblock


\bibitem[Labutov et~al\mbox{.}(2018)]%
        {labutov-etal-2018-learning}
\bibfield{author}{\bibinfo{person}{Igor Labutov}, \bibinfo{person}{Bishan Yang}, {and} \bibinfo{person}{Tom Mitchell}.} \bibinfo{year}{2018}\natexlab{}.
\newblock \showarticletitle{Learning to Learn Semantic Parsers from Natural Language Supervision}. In \bibinfo{booktitle}{\emph{Proceedings of the 2018 Conference on Empirical Methods in Natural Language Processing}}, \bibfield{editor}{\bibinfo{person}{Ellen Riloff}, \bibinfo{person}{David Chiang}, \bibinfo{person}{Julia Hockenmaier}, {and} \bibinfo{person}{Jun{'}ichi Tsujii}} (Eds.). \bibinfo{publisher}{Association for Computational Linguistics}, \bibinfo{address}{Brussels, Belgium}, \bibinfo{pages}{1676--1690}.
\newblock
\urldef\tempurl%
\url{https://doi.org/10.18653/v1/D18-1195}
\showDOI{\tempurl}


\bibitem[Leinonen et~al\mbox{.}(2023)]%
        {leinonen2023comparing}
\bibfield{author}{\bibinfo{person}{Juho Leinonen}, \bibinfo{person}{Paul Denny}, \bibinfo{person}{Stephen MacNeil}, \bibinfo{person}{Sami Sarsa}, \bibinfo{person}{Seth Bernstein}, \bibinfo{person}{Joanne Kim}, \bibinfo{person}{Andrew Tran}, {and} \bibinfo{person}{Arto Hellas}.} \bibinfo{year}{2023}\natexlab{}.
\newblock \showarticletitle{Comparing code explanations created by students and large language models}. In \bibinfo{booktitle}{\emph{Proceedings of the 2023 Conference on Innovation and Technology in Computer Science Education V. 1}}. \bibinfo{pages}{124--130}.
\newblock


\bibitem[Li et~al\mbox{.}(2023)]%
        {li2023starcoder}
\bibfield{author}{\bibinfo{person}{Raymond Li}, \bibinfo{person}{Loubna~Ben Allal}, \bibinfo{person}{Yangtian Zi}, \bibinfo{person}{Niklas Muennighoff}, \bibinfo{person}{Denis Kocetkov}, \bibinfo{person}{Chenghao Mou}, \bibinfo{person}{Marc Marone}, \bibinfo{person}{Christopher Akiki}, \bibinfo{person}{Jia Li}, \bibinfo{person}{Jenny Chim}, {et~al\mbox{.}}} \bibinfo{year}{2023}\natexlab{}.
\newblock \showarticletitle{StarCoder: may the source be with you!}
\newblock \bibinfo{journal}{\emph{arXiv preprint arXiv:2305.06161}} (\bibinfo{year}{2023}).
\newblock


\bibitem[Li et~al\mbox{.}(2020)]%
        {li-etal-2020-mean}
\bibfield{author}{\bibinfo{person}{Yuntao Li}, \bibinfo{person}{Bei Chen}, \bibinfo{person}{Qian Liu}, \bibinfo{person}{Yan Gao}, \bibinfo{person}{Jian-Guang Lou}, \bibinfo{person}{Yan Zhang}, {and} \bibinfo{person}{Dongmei Zhang}.} \bibinfo{year}{2020}\natexlab{}.
\newblock \showarticletitle{{``}What Do You Mean by That?{''} A Parser-Independent Interactive Approach for Enhancing Text-to-{SQL}}. In \bibinfo{booktitle}{\emph{Proceedings of the 2020 Conference on Empirical Methods in Natural Language Processing (EMNLP)}}, \bibfield{editor}{\bibinfo{person}{Bonnie Webber}, \bibinfo{person}{Trevor Cohn}, \bibinfo{person}{Yulan He}, {and} \bibinfo{person}{Yang Liu}} (Eds.). \bibinfo{publisher}{Association for Computational Linguistics}, \bibinfo{address}{Online}, \bibinfo{pages}{6913--6922}.
\newblock
\urldef\tempurl%
\url{https://doi.org/10.18653/v1/2020.emnlp-main.561}
\showDOI{\tempurl}


\bibitem[Liu et~al\mbox{.}(2023)]%
        {liu2023pre}
\bibfield{author}{\bibinfo{person}{Pengfei Liu}, \bibinfo{person}{Weizhe Yuan}, \bibinfo{person}{Jinlan Fu}, \bibinfo{person}{Zhengbao Jiang}, \bibinfo{person}{Hiroaki Hayashi}, {and} \bibinfo{person}{Graham Neubig}.} \bibinfo{year}{2023}\natexlab{}.
\newblock \showarticletitle{Pre-train, prompt, and predict: A systematic survey of prompting methods in natural language processing}.
\newblock \bibinfo{journal}{\emph{Comput. Surveys}} \bibinfo{volume}{55}, \bibinfo{number}{9} (\bibinfo{year}{2023}), \bibinfo{pages}{1--35}.
\newblock


\bibitem[Mo et~al\mbox{.}(2022)]%
        {mo-etal-2022-towards}
\bibfield{author}{\bibinfo{person}{Lingbo Mo}, \bibinfo{person}{Ashley Lewis}, \bibinfo{person}{Huan Sun}, {and} \bibinfo{person}{Michael White}.} \bibinfo{year}{2022}\natexlab{}.
\newblock \showarticletitle{Towards Transparent Interactive Semantic Parsing via Step-by-Step Correction}. In \bibinfo{booktitle}{\emph{Findings of the Association for Computational Linguistics: ACL 2022}}, \bibfield{editor}{\bibinfo{person}{Smaranda Muresan}, \bibinfo{person}{Preslav Nakov}, {and} \bibinfo{person}{Aline Villavicencio}} (Eds.). \bibinfo{publisher}{Association for Computational Linguistics}, \bibinfo{address}{Dublin, Ireland}, \bibinfo{pages}{322--342}.
\newblock
\urldef\tempurl%
\url{https://doi.org/10.18653/v1/2022.findings-acl.28}
\showDOI{\tempurl}


\bibitem[Nam et~al\mbox{.}(2024)]%
        {nam2024using}
\bibfield{author}{\bibinfo{person}{Daye Nam}, \bibinfo{person}{Andrew Macvean}, \bibinfo{person}{Vincent Hellendoorn}, \bibinfo{person}{Bogdan Vasilescu}, {and} \bibinfo{person}{Brad Myers}.} \bibinfo{year}{2024}\natexlab{}.
\newblock \showarticletitle{Using an llm to help with code understanding}. In \bibinfo{booktitle}{\emph{Proceedings of the IEEE/ACM 46th International Conference on Software Engineering}}. \bibinfo{pages}{1--13}.
\newblock


\bibitem[OpenAI(2023)]%
        {openai:gpt}
\bibfield{author}{\bibinfo{person}{OpenAI}.} \bibinfo{year}{2023}\natexlab{}.
\newblock \bibinfo{title}{ChatGPT}.
\newblock \bibinfo{howpublished}{\url{https://openai.com}}.
\newblock


\bibitem[Prather et~al\mbox{.}(2023)]%
        {prather2023s}
\bibfield{author}{\bibinfo{person}{James Prather}, \bibinfo{person}{Brent~N Reeves}, \bibinfo{person}{Paul Denny}, \bibinfo{person}{Brett~A Becker}, \bibinfo{person}{Juho Leinonen}, \bibinfo{person}{Andrew Luxton-Reilly}, \bibinfo{person}{Garrett Powell}, \bibinfo{person}{James Finnie-Ansley}, {and} \bibinfo{person}{Eddie~Antonio Santos}.} \bibinfo{year}{2023}\natexlab{}.
\newblock \showarticletitle{“It’s Weird That it Knows What I Want”: Usability and Interactions with Copilot for Novice Programmers}.
\newblock \bibinfo{journal}{\emph{ACM Transactions on Computer-Human Interaction}} \bibinfo{volume}{31}, \bibinfo{number}{1} (\bibinfo{year}{2023}), \bibinfo{pages}{1--31}.
\newblock


\bibitem[Rabbi et~al\mbox{.}(2024)]%
        {rabbi2024ai}
\bibfield{author}{\bibinfo{person}{Md~Fazle Rabbi}, \bibinfo{person}{Arifa~Islam Champa}, \bibinfo{person}{Minhaz~F Zibran}, {and} \bibinfo{person}{Md~Rakibul Islam}.} \bibinfo{year}{2024}\natexlab{}.
\newblock \showarticletitle{AI writes, we analyze: The ChatGPT python code saga}. In \bibinfo{booktitle}{\emph{Proceedings of the 21st International Conference on Mining Software Repositories}}. \bibinfo{pages}{177--181}.
\newblock


\bibitem[Ross et~al\mbox{.}(2023)]%
        {ross2023programmer}
\bibfield{author}{\bibinfo{person}{Steven~I Ross}, \bibinfo{person}{Fernando Martinez}, \bibinfo{person}{Stephanie Houde}, \bibinfo{person}{Michael Muller}, {and} \bibinfo{person}{Justin~D Weisz}.} \bibinfo{year}{2023}\natexlab{}.
\newblock \showarticletitle{The programmer’s assistant: Conversational interaction with a large language model for software development}. In \bibinfo{booktitle}{\emph{Proceedings of the 28th International Conference on Intelligent User Interfaces}}. \bibinfo{pages}{491--514}.
\newblock


\bibitem[Roziere et~al\mbox{.}(2023)]%
        {roziere2023code}
\bibfield{author}{\bibinfo{person}{Baptiste Roziere}, \bibinfo{person}{Jonas Gehring}, \bibinfo{person}{Fabian Gloeckle}, \bibinfo{person}{Sten Sootla}, \bibinfo{person}{Itai Gat}, \bibinfo{person}{Xiaoqing~Ellen Tan}, \bibinfo{person}{Yossi Adi}, \bibinfo{person}{Jingyu Liu}, \bibinfo{person}{Tal Remez}, \bibinfo{person}{J{\'e}r{\'e}my Rapin}, {et~al\mbox{.}}} \bibinfo{year}{2023}\natexlab{}.
\newblock \showarticletitle{Code llama: Open foundation models for code}.
\newblock \bibinfo{journal}{\emph{arXiv preprint arXiv:2308.12950}} (\bibinfo{year}{2023}).
\newblock


\bibitem[Sarsa et~al\mbox{.}(2022)]%
        {sarsa2022automatic}
\bibfield{author}{\bibinfo{person}{Sami Sarsa}, \bibinfo{person}{Paul Denny}, \bibinfo{person}{Arto Hellas}, {and} \bibinfo{person}{Juho Leinonen}.} \bibinfo{year}{2022}\natexlab{}.
\newblock \showarticletitle{Automatic generation of programming exercises and code explanations using large language models}. In \bibinfo{booktitle}{\emph{Proceedings of the 2022 ACM Conference on International Computing Education Research-Volume 1}}. \bibinfo{pages}{27--43}.
\newblock


\bibitem[Sheese et~al\mbox{.}(2024)]%
        {sheese2024patterns}
\bibfield{author}{\bibinfo{person}{Brad Sheese}, \bibinfo{person}{Mark Liffiton}, \bibinfo{person}{Jaromir Savelka}, {and} \bibinfo{person}{Paul Denny}.} \bibinfo{year}{2024}\natexlab{}.
\newblock \showarticletitle{Patterns of student help-seeking when using a large language model-powered programming assistant}. In \bibinfo{booktitle}{\emph{Proceedings of the 26th Australasian Computing Education Conference}}. \bibinfo{pages}{49--57}.
\newblock


\bibitem[Shin(2019)]%
        {shin2019encoding}
\bibfield{author}{\bibinfo{person}{Richard Shin}.} \bibinfo{year}{2019}\natexlab{}.
\newblock \showarticletitle{Encoding database schemas with relation-aware self-attention for text-to-sql parsers}.
\newblock \bibinfo{journal}{\emph{arXiv preprint arXiv:1906.11790}} (\bibinfo{year}{2019}).
\newblock


\bibitem[Sridhara et~al\mbox{.}(2023)]%
        {sridhara2023chatgpt}
\bibfield{author}{\bibinfo{person}{Giriprasad Sridhara}, \bibinfo{person}{Sourav Mazumdar}, {et~al\mbox{.}}} \bibinfo{year}{2023}\natexlab{}.
\newblock \showarticletitle{Chatgpt: A study on its utility for ubiquitous software engineering tasks}.
\newblock \bibinfo{journal}{\emph{arXiv preprint arXiv:2305.16837}} (\bibinfo{year}{2023}).
\newblock


\bibitem[Staniek and Riezler(2021)]%
        {staniek2021error}
\bibfield{author}{\bibinfo{person}{Michael Staniek} {and} \bibinfo{person}{Stefan Riezler}.} \bibinfo{year}{2021}\natexlab{}.
\newblock \showarticletitle{Error-Aware Interactive Semantic Parsing of OpenStreetMap}. In \bibinfo{booktitle}{\emph{Proceedings of Second International Combined Workshop on Spatial Language Understanding and Grounded Communication for Robotics}}. \bibinfo{pages}{53--59}.
\newblock


\bibitem[Su et~al\mbox{.}(2018)]%
        {su2018natural}
\bibfield{author}{\bibinfo{person}{Yu Su}, \bibinfo{person}{Ahmed Hassan~Awadallah}, \bibinfo{person}{Miaosen Wang}, {and} \bibinfo{person}{Ryen~W White}.} \bibinfo{year}{2018}\natexlab{}.
\newblock \showarticletitle{Natural language interfaces with fine-grained user interaction: A case study on web apis}. In \bibinfo{booktitle}{\emph{The 41st International ACM SIGIR Conference on Research \& Development in Information Retrieval}}. \bibinfo{pages}{855--864}.
\newblock


\bibitem[Surameery and Shakor(2023)]%
        {surameery2023use}
\bibfield{author}{\bibinfo{person}{Nigar M~Shafiq Surameery} {and} \bibinfo{person}{Mohammed~Y Shakor}.} \bibinfo{year}{2023}\natexlab{}.
\newblock \showarticletitle{Use chat gpt to solve programming bugs}.
\newblock \bibinfo{journal}{\emph{International Journal of Information Technology and Computer Engineering}} \bibinfo{number}{31} (\bibinfo{year}{2023}), \bibinfo{pages}{17--22}.
\newblock


\bibitem[Vaithilingam et~al\mbox{.}(2022)]%
        {vaithilingam2022expectation}
\bibfield{author}{\bibinfo{person}{Priyan Vaithilingam}, \bibinfo{person}{Tianyi Zhang}, {and} \bibinfo{person}{Elena~L Glassman}.} \bibinfo{year}{2022}\natexlab{}.
\newblock \showarticletitle{Expectation vs. experience: Evaluating the usability of code generation tools powered by large language models}. In \bibinfo{booktitle}{\emph{Chi conference on human factors in computing systems extended abstracts}}. \bibinfo{pages}{1--7}.
\newblock


\bibitem[Wang et~al\mbox{.}(2023)]%
        {wang2023leti}
\bibfield{author}{\bibinfo{person}{Xingyao Wang}, \bibinfo{person}{Hao Peng}, \bibinfo{person}{Reyhaneh Jabbarvand}, {and} \bibinfo{person}{Heng Ji}.} \bibinfo{year}{2023}\natexlab{}.
\newblock \showarticletitle{LeTI: Learning to Generate from Textual Interactions}.
\newblock \bibinfo{journal}{\emph{arXiv preprint arXiv:2305.10314}} (\bibinfo{year}{2023}).
\newblock


\bibitem[Wang et~al\mbox{.}(2022)]%
        {wang2022compilable}
\bibfield{author}{\bibinfo{person}{Xin Wang}, \bibinfo{person}{Yasheng Wang}, \bibinfo{person}{Yao Wan}, \bibinfo{person}{Fei Mi}, \bibinfo{person}{Yitong Li}, \bibinfo{person}{Pingyi Zhou}, \bibinfo{person}{Jin Liu}, \bibinfo{person}{Hao Wu}, \bibinfo{person}{Xin Jiang}, {and} \bibinfo{person}{Qun Liu}.} \bibinfo{year}{2022}\natexlab{}.
\newblock \showarticletitle{Compilable Neural Code Generation with Compiler Feedback}. In \bibinfo{booktitle}{\emph{Findings of the Association for Computational Linguistics: ACL 2022}}. \bibinfo{pages}{9--19}.
\newblock


\bibitem[Wang et~al\mbox{.}(2021)]%
        {wang2021codet5}
\bibfield{author}{\bibinfo{person}{Yue Wang}, \bibinfo{person}{Weishi Wang}, \bibinfo{person}{Shafiq Joty}, {and} \bibinfo{person}{Steven~CH Hoi}.} \bibinfo{year}{2021}\natexlab{}.
\newblock \showarticletitle{Codet5: Identifier-aware unified pre-trained encoder-decoder models for code understanding and generation}.
\newblock \bibinfo{journal}{\emph{arXiv preprint arXiv:2109.00859}} (\bibinfo{year}{2021}).
\newblock


\bibitem[Xiao et~al\mbox{.}(2024)]%
        {xiao2024devgpt}
\bibfield{author}{\bibinfo{person}{Tao Xiao}, \bibinfo{person}{Christoph Treude}, \bibinfo{person}{Hideaki Hata}, {and} \bibinfo{person}{Kenichi Matsumoto}.} \bibinfo{year}{2024}\natexlab{}.
\newblock \showarticletitle{Devgpt: Studying developer-chatgpt conversations}. In \bibinfo{booktitle}{\emph{2024 IEEE/ACM 21st International Conference on Mining Software Repositories (MSR)}}. IEEE, \bibinfo{pages}{227--230}.
\newblock


\bibitem[Yan et~al\mbox{.}(2024)]%
        {yan2024ivie}
\bibfield{author}{\bibinfo{person}{Litao Yan}, \bibinfo{person}{Alyssa Hwang}, \bibinfo{person}{Zhiyuan Wu}, {and} \bibinfo{person}{Andrew Head}.} \bibinfo{year}{2024}\natexlab{}.
\newblock \showarticletitle{Ivie: Lightweight anchored explanations of just-generated code}. In \bibinfo{booktitle}{\emph{Proceedings of the CHI Conference on Human Factors in Computing Systems}}. \bibinfo{pages}{1--15}.
\newblock


\bibitem[Yang et~al\mbox{.}(2023)]%
        {yang2023intercode}
\bibfield{author}{\bibinfo{person}{John Yang}, \bibinfo{person}{Akshara Prabhakar}, \bibinfo{person}{Karthik Narasimhan}, {and} \bibinfo{person}{Shunyu Yao}.} \bibinfo{year}{2023}\natexlab{}.
\newblock \showarticletitle{InterCode: Standardizing and Benchmarking Interactive Coding with Execution Feedback}.
\newblock \bibinfo{journal}{\emph{arXiv preprint arXiv:2306.14898}} (\bibinfo{year}{2023}).
\newblock


\bibitem[Yao et~al\mbox{.}(2019a)]%
        {yao2019interactive}
\bibfield{author}{\bibinfo{person}{Ziyu Yao}, \bibinfo{person}{Xiujun Li}, \bibinfo{person}{Jianfeng Gao}, \bibinfo{person}{Brian Sadler}, {and} \bibinfo{person}{Huan Sun}.} \bibinfo{year}{2019}\natexlab{a}.
\newblock \showarticletitle{Interactive semantic parsing for if-then recipes via hierarchical reinforcement learning}. In \bibinfo{booktitle}{\emph{Proceedings of the AAAI Conference on Artificial Intelligence}}, Vol.~\bibinfo{volume}{33}. \bibinfo{pages}{2547--2554}.
\newblock


\bibitem[Yao et~al\mbox{.}(2019b)]%
        {yao-etal-2019-model}
\bibfield{author}{\bibinfo{person}{Ziyu Yao}, \bibinfo{person}{Yu Su}, \bibinfo{person}{Huan Sun}, {and} \bibinfo{person}{Wen-tau Yih}.} \bibinfo{year}{2019}\natexlab{b}.
\newblock \showarticletitle{Model-based Interactive Semantic Parsing: A Unified Framework and A Text-to-{SQL} Case Study}. In \bibinfo{booktitle}{\emph{Proceedings of the 2019 Conference on Empirical Methods in Natural Language Processing and the 9th International Joint Conference on Natural Language Processing (EMNLP-IJCNLP)}}, \bibfield{editor}{\bibinfo{person}{Kentaro Inui}, \bibinfo{person}{Jing Jiang}, \bibinfo{person}{Vincent Ng}, {and} \bibinfo{person}{Xiaojun Wan}} (Eds.). \bibinfo{publisher}{Association for Computational Linguistics}, \bibinfo{address}{Hong Kong, China}, \bibinfo{pages}{5447--5458}.
\newblock
\urldef\tempurl%
\url{https://doi.org/10.18653/v1/D19-1547}
\showDOI{\tempurl}


\bibitem[Yao et~al\mbox{.}(2020)]%
        {yao-etal-2020-imitation}
\bibfield{author}{\bibinfo{person}{Ziyu Yao}, \bibinfo{person}{Yiqi Tang}, \bibinfo{person}{Wen-tau Yih}, \bibinfo{person}{Huan Sun}, {and} \bibinfo{person}{Yu Su}.} \bibinfo{year}{2020}\natexlab{}.
\newblock \showarticletitle{An Imitation Game for Learning Semantic Parsers from User Interaction}. In \bibinfo{booktitle}{\emph{Proceedings of the 2020 Conference on Empirical Methods in Natural Language Processing (EMNLP)}}, \bibfield{editor}{\bibinfo{person}{Bonnie Webber}, \bibinfo{person}{Trevor Cohn}, \bibinfo{person}{Yulan He}, {and} \bibinfo{person}{Yang Liu}} (Eds.). \bibinfo{publisher}{Association for Computational Linguistics}, \bibinfo{address}{Online}, \bibinfo{pages}{6883--6902}.
\newblock
\urldef\tempurl%
\url{https://doi.org/10.18653/v1/2020.emnlp-main.559}
\showDOI{\tempurl}


\bibitem[Yu et~al\mbox{.}(2018)]%
        {yu-etal-2018-spider}
\bibfield{author}{\bibinfo{person}{Tao Yu}, \bibinfo{person}{Rui Zhang}, \bibinfo{person}{Kai Yang}, \bibinfo{person}{Michihiro Yasunaga}, \bibinfo{person}{Dongxu Wang}, \bibinfo{person}{Zifan Li}, \bibinfo{person}{James Ma}, \bibinfo{person}{Irene Li}, \bibinfo{person}{Qingning Yao}, \bibinfo{person}{Shanelle Roman}, \bibinfo{person}{Zilin Zhang}, {and} \bibinfo{person}{Dragomir Radev}.} \bibinfo{year}{2018}\natexlab{}.
\newblock \showarticletitle{{S}pider: A Large-Scale Human-Labeled Dataset for Complex and Cross-Domain Semantic Parsing and Text-to-{SQL} Task}. In \bibinfo{booktitle}{\emph{Proceedings of the 2018 Conference on Empirical Methods in Natural Language Processing}}, \bibfield{editor}{\bibinfo{person}{Ellen Riloff}, \bibinfo{person}{David Chiang}, \bibinfo{person}{Julia Hockenmaier}, {and} \bibinfo{person}{Jun{'}ichi Tsujii}} (Eds.). \bibinfo{publisher}{Association for Computational Linguistics}, \bibinfo{address}{Brussels, Belgium}, \bibinfo{pages}{3911--3921}.
\newblock
\urldef\tempurl%
\url{https://doi.org/10.18653/v1/D18-1425}
\showDOI{\tempurl}


\bibitem[Zhang et~al\mbox{.}(2020)]%
        {zhang2020interactive}
\bibfield{author}{\bibinfo{person}{Tianyi Zhang}, \bibinfo{person}{London Lowmanstone}, \bibinfo{person}{Xinyu Wang}, {and} \bibinfo{person}{Elena~L Glassman}.} \bibinfo{year}{2020}\natexlab{}.
\newblock \showarticletitle{Interactive program synthesis by augmented examples}. In \bibinfo{booktitle}{\emph{Proceedings of the 33rd Annual ACM Symposium on User Interface Software and Technology}}. \bibinfo{pages}{627--648}.
\newblock


\end{thebibliography}


\end{document}